\documentclass[11pt]{article}
\usepackage{a4wide}
\usepackage{amsfonts} 
\usepackage{psfrag}
\usepackage{graphics}
\usepackage{subfigure}
\usepackage{ifpdf}
\usepackage{epsf}
\usepackage{color}
\ifpdf
\usepackage{epstopdf}
\usepackage{hyperref}
\else
\usepackage[hypertex]{hyperref}
\fi

\newcommand{\eref}[1]{(\ref{#1})}

\newcommand{\tfrac}[2]{{\textstyle \frac{#1}{#2}}}
\newcommand{\blank}[1]{}

\newcommand\sect[1]{\section{#1}\setcounter{equation}0} 

\newcommand\void[1]       {}

\newcommand\be            {\begin{equation}}
\newcommand\bea           {\begin{eqnarray}}
\newcommand\rd             {{\mathrm d}}
\newcommand\ee            {\end{equation}}
\newcommand\eea           {\end{eqnarray}}

\newcommand\id            {{\rm id}}

\newcommand\Hc            {\mathcal{H}}
\newcommand\cH            {\mathcal{H}}

\newcommand\Nc            {\mathcal{N}}

\newcommand\Pc            {\mathcal{P}}

\renewcommand\vec[1]{{\vert{#1}\rangle}}

\newcommand\cev[1]{{\langle{#1}\vert}}

\newcommand\vac{{\vec 0}}

\def\3pt#1#2#3{{\langle{#1}|{#2}|{#3}\rangle}}

\def\cbI#1#2#3#4#5#6#7#8{
\setlength{\unitlength}{#1sp}%
\centering{
\begin{picture}(1500,757)(4500,-2483)
\thicklines
{\put(4850,-1860){\line( 1,0){800}}}%
{\put(4850,-2460){\line( 1,0){800}}}%
{\put(5250,-1860){\line( 0,-1){600}}}%
\put(4775,-2460){\makebox(0,0)[rc]{$#2$}}
\put(4775,-1860){\makebox(0,0)[rc]{$#3$}}
\put(5750,-1860){\makebox(0,0)[lc]{$#4$}}
\put(5750,-2460){\makebox(0,0)[lc]{$#5$}}
\put(5300,-2160){\makebox(0,0)[lc]{$#6$}}
\end{picture}}}%

\def\cbb#1#2#3#4#5#6#7#8{
\setlength{\unitlength}{#1sp}%
\begin{picture}(2800,1400)(3901,-2683)
\thicklines
{\put(4800,-1860){\line( 0,-1){600}}}%
{\put(5700,-1860){\line( 0,-1){600}}}%
{\put(4200,-2460){\line( 1, 0){2100}}}%
\put(4150,-2460){\makebox(0,0)[rc]{$#2$}}
\put(4800,-1800){\makebox(0,0)[cb]{$#3$}}
\put(4800,-2560){\makebox(0,0)[ct]{$#7$}}
\put(5250,-2380){\makebox(0,0)[cb]{$#4$}}
\put(5700,-1800){\makebox(0,0)[cb]{$#5$}}
\put(5700,-2560){\makebox(0,0)[ct]{$#8$}}
\put(6400,-2460){\makebox(0,0)[lc]{$#6$}}
\end{picture}}%

\def\thefootnote{\fnsymbol{footnote}}

\begin{document}

\begin{flushright}  {~} \\[-12mm]
{\sf KCL-MTH-11-04}\\[1mm]
\end{flushright} 

\thispagestyle{empty}

\begin{center} \vskip 14mm
{\Large\bf On the renormalisation group for the boundary Truncated
  Conformal Space Approach}\\[20mm] 
{\large 
G\'erard M.\ T.\ Watts~~\footnote{Email: gerard.watts@kcl.ac.uk}}
\\[8mm]
Department of Mathematics, King's College London,\\
Strand, London WC2R 2LS -- UK

\vskip 22mm
\end{center}

\begin{quote}{\bf Abstract}\\[1mm]
In this paper we continue the study of the truncated conformal space
approach to perturbed boundary conformal field theories. This approach
to perturbation theory suffers from a renormalisation of the coupling
constant and a multiplicative 
renormalisation of the Hamiltonian. We show how these two effects can
be predicted by both physical and mathematical arguments and prove
that they are correct to leading order for all states in the TCSA
system. We check these results using the TCSA applied to the
tri-critical Ising model and the Yang-Lee model.
We also study the TCSA of an irrelevant (non-renormalisable)
perturbation and find that, while the convergence of the coupling
constant and energy scales are problematic, the renormalised and
rescaled spectrum remain a very good fit to the exact result, and we
find a numerical relationship between the IR and UV couplings
describing a particular flow.
Finally we study the large coupling behaviour of TCSA and show that it
accurately encompasses several different fixed points.
\end{quote}

\vfill
\newpage 

\setcounter{footnote}{0}
\def\thefootnote{\arabic{footnote}}

\sect{Introduction}

The Truncated Conformal Space approach (TCSA) of Yurov and
Zamolodchikov \cite{YZ1} is a widely-used method
to study the finite-size dependence of perturbed two-dimensional
conformal field theories. It is based on truncating the infinite
dimensional Hilbert space to a finite-dimensional system on which the
Hamiltonian is studied numerically.\footnote{%
In this paper we study the original form due to Yurov and
Zamolodchikov, not the revised version of
\cite{K1,K2}. 
}
It has been known for a long time
that the method has various convergence problems which can reduce its
effectiveness
\cite{LCM1}. 
The principal problems are a renormalisation of the
coupling constant, a renormalisation of the energy scale and
differences between ground state contributions in different sectors,
all of which depend on the size of the truncated system (we shall
always cut off the size of system by taking all states whose unperturbed
energy above the ground state is less than or equal to a given number,
which we shall call the truncation level). In a 
previous paper we 
showed how the renormalisation of the coupling constant in perturbed
boundary conformal field theories could be studied using a variant of
standard perturbed-conformal field theory methods \cite{FGPTW}. In
this paper we 
extend this study to the second effect, and show how the leading energy
scale renormalisation is an overall multiplicative renormalisation
which can be considered equivalent to a renormalisation of the size of
the system. We find `physical' arguments based on the operator
product expansions and also more rigorous arguments based on an
analysis of the eigenvalues of the perturbed Hamiltonian and show that
these give identical results. We test our
results using two integrable conformal field theories, the
tri-critical Ising model and the Yang-Lee model, as in both cases the
finite-size spectrum has been found using TBA methods and this
provides an accurate quantitative check of the proposed results.

In the next section we briefly review the case of the tri-critical
model to illustrate the renormalisation issues to be solved and to
find numerical estimates for the coupling renormalisation and energy
rescaling. We then show that these have scaling forms and find
numerical estimates for the associated exponents. 

In the third section we show how these can be derived, to one loop
order, from considerations of the renormalisation of the perturbed
action and in the fourth section show how these can be proven, to
leading order, for all energy levels, from analysis of the
eigenvalues of the perturbed Hamiltonian.
In the subsequent two sections we check these predictions against the
numerical data in the tri-critical Ising model and the Yang-Lee model

In section \ref{sec:irrel} we consider the case of an irrelevant
(non-renormalisable) perturbation and 
in section \ref{sec:reverse} we consider the flows beyond the fixed
points and speculate on the exponents that have been found
numerically. Finally in section \ref{sec:conc} we present our
conclusions.

\sect{The TCSA approach and its errors}

\subsection{The TCSA approach}

We start with a CFT defined on a strip $0\leq y\leq L$ of width $L$
in the upper half 
plane with coordinate $z=x+iy$. We take the strip to have conformally
invariant boundary conditions so that the system is conformally
invariant, the Hilbert space carries an action of the Virasoro
algebra and decomposes into a direct sum of representations of the
Virasoro algebra. The representations occurring depend on the boundary
conditions on the strip.

The unperturbed CFT Hamiltonian generating translations along the
strip is
\be
 H = \int_0^L T_{xx}\;\frac{\rd y}{2\pi}
\;.
\label{eq:Hcft}
\ee
We will map the strip to the upper half plane with coordinate
$w=\exp(\pi z/L)$ in terms of which the CFT Hamiltonian is 
\be
 H = \frac{\pi}{L}\left( L_0 - \frac{c}{24} \right)
\;,
\label{eq:Hcft2}
\ee
where $L_0$ is the zero mode of the Virasoro algebra.

We are interested in perturbations by one or more boundary fields
$\phi_i(x)$ living on the bottom edge of the strip, $y=0$. We take
these to be quasi-primary fields of conformal dimension $h_i$.
If the coupling to these fields are $\mu_i$ then the perturbation is
given by an addition to the action 
\be
\delta S = \int \sum_i \mu_i \phi_i(x)\,\rd x
\;.
\ee
When mapped to the upper half plane this gives the perturbed
Hamiltonian as
\be
 H = \frac{\pi}{L}
   \left[\;
   \left( L_0 - \frac{c}{24} \right)
  \;+\; \sum_i\mu_i\left(\frac{L}{\pi}\right)^{y_i}\,\phi_i(1)
   \right]
\;,
\label{eq:Hpcft1}
\ee
where $y_i = 1- h_i$. Note that 
\eref{eq:Hpcft1} is only correct if the fields $\phi_i$ are primary;
there are corrections if they are quasi-primary but not primary.
We will normally consider the dimensionless operator
\be
 \Hc
= \left(\frac{L}{\pi}\right) H
=  \left( L_0 - \frac{c}{24} \right)
  \;+\; \sum_i\frac{\lambda_i}{\pi^{y_i}}\,\phi_i(1)
  \;,
\label{eq:Hpcft2}
\ee
where $\lambda_i = \mu_i L^y$ are dimensionless coupling constants.

The TCSA approach is to restrict this Hamiltonian to a finite
dimensional space of excitation level $n$ or lower. If we take the
projector onto this space to be $P_n$ then the TCSA Hamiltonians are
\be
 \Hc_n
=  \left( L_0 - \frac{c}{24} \right)
  \;+\; 
  \sum_i \frac{\lambda_i}{\pi^{y_i}}\,P_n\phi_i(1)P_n
  \;.
\label{eq:Hpcft3}
\ee
These Hamiltonians can be diagonalised numerically and their eigenvalues
and eigenstates form the TCSA
approximations to the perturbed system. Any quantity that can be
considered in the perturbed system can also be considered in the TCSA
system; the only question is how good the approximation is, and
whether the dependence on the truncation level is either small or can
be estimated efficiently. In the next section we present the two
leading truncation effects --- the renormalisation of the coupling
constant and the rescaling of the energy levels --- in the case of the
tri-critical Ising model.

\subsection{The tri-critical Ising model numerical results}

We take as our example model example the boundary tri-critical Ising model on
a strip. 
We shall review the boundary conditions and their flows in section
\ref{sec:tcimcheck}; for the moment it is sufficient to know that,
amongst others,
there are conformal boundary conditions labelled $(11)$, $(21)$ and $(12)$,
and the $(12)$ boundary condition can be perturbed 
by a field of weight $3/5$ with the following flows:
\be
 (11) \longleftarrow (12) \longrightarrow (12)
\;.
\label{eq:basicflow}
\ee
\textmd{}The spectrum of the Hamiltonian on a strip with an unperturbed $(11)$
and a perturbed $(12)$ boundary condition has also been 
analysed from the TBA approach by Feverati and
collaborators in \cite{FPR1}. Using the results in 
\cite{NepomechieAhn} and \cite{BLZ1,BLZ2}, we can relate the TBA and
conformal perturbation theory parameters and 
so we can compare the TCSA spectrum to the exact perturbed conformal
field theory spectrum for this system. 

\begin{figure}[thb]
\subfigure[TBA data]{\scalebox{0.7}{\includegraphics{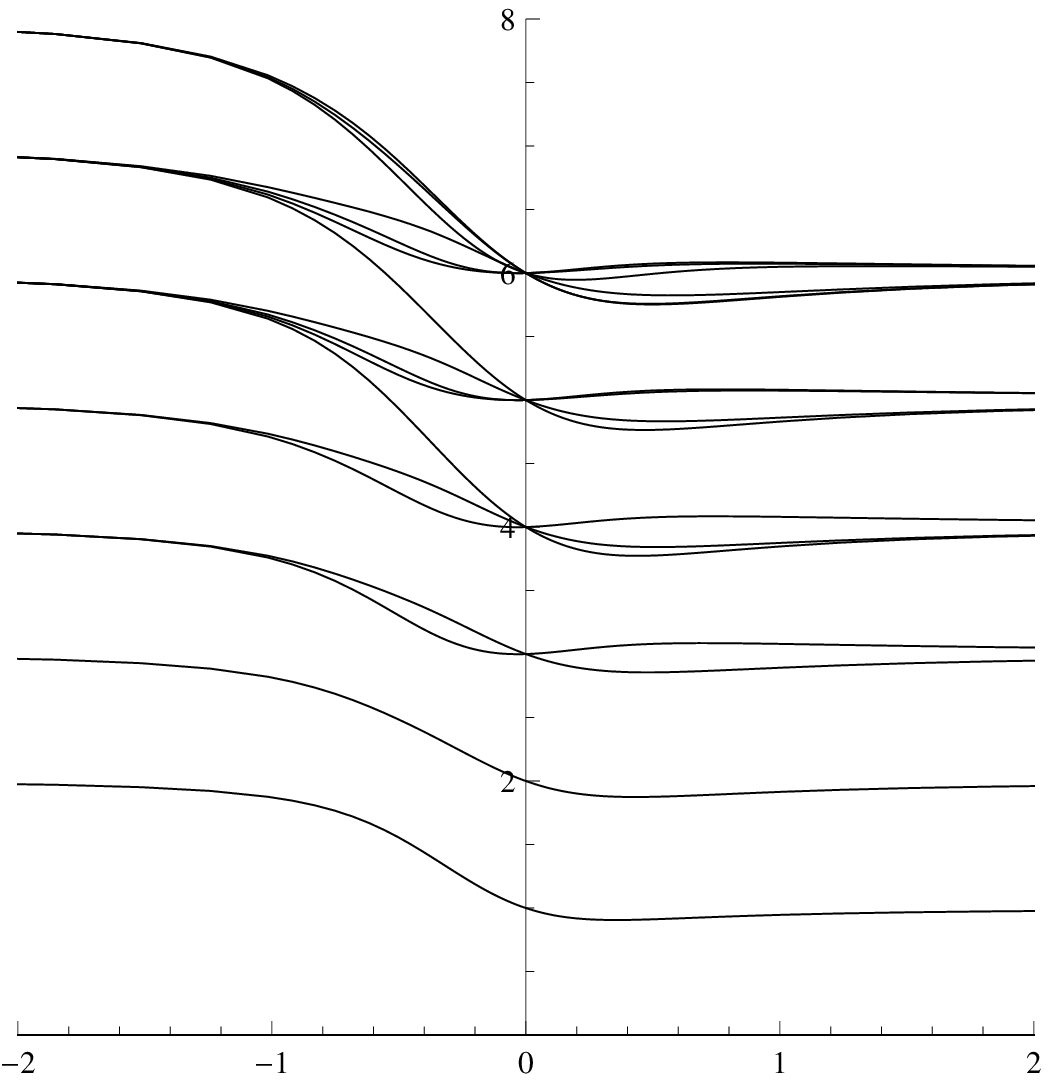}}
\label{fig:tbaflows}}
\hfill
\subfigure[TCSA data from truncation level 14]%
{\scalebox{0.7}{\includegraphics{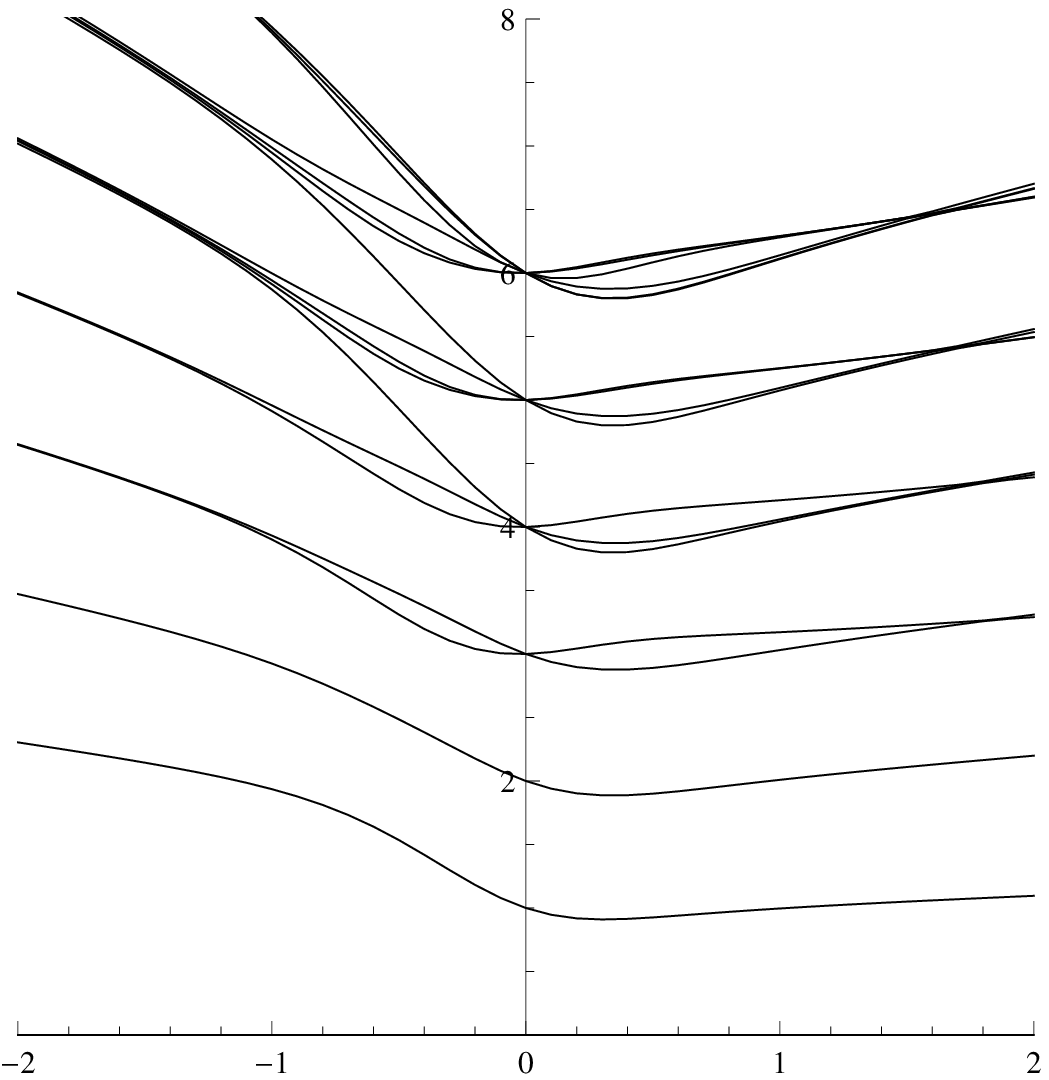}}%
\label{fig:tcsaflows}}
\caption{The energy gaps for the strip
  with boundary conditions $(11)$ and $(12) + \lambda\phi_{(13)}$ as
  given by the TBA and TCSA methods plotted against $\lambda$.} 
\label{figure3}
\end{figure}

The most immediate difference is that there is an overall shift in the
energy levels. For finite perturbations for which there are no
divergences in the perturbation expansion this corresponds to a
measurable free energy per unit length. In this case the perturbation
theory is divergent and the overall shift in the energies is not
meaningful: in TCSA it depends on the truncation level and does not
converge to a finite value.
As a consequence, in figure \ref{figure3} we 
just show the difference between the ground state and the
excited states - the energy gaps: in figure
\ref{fig:tbaflows} we give the gaps as calculated using the TBA
method, and in figure \ref{fig:tcsaflows} the TCSA gaps. 

On inspection of figure \ref{figure3}, one can also discern the two
other effects. In 
the TCSA plot, there is an apparent convergence of many levels at or
near $\lambda=2$. This corresponds to a fixed point of the flow where
the $(12)$ boundary condition has reached its $(21)$ fixed point. This
is only reached at $\lambda=\infty$ in the TBA flow. The two couplings
are related by a non-trivial renormalisation.  Less obvious but still
measurable is the overall rescaling of the energy levels. At
$\lambda=-2$ the TCSA spectrum has almost regrouped into
equally-spaced levels but the spacing is approximately
15\% larger than the (correct) spacing in the TBA spectrum. This
effect grows markedly with increasing $|\lambda|$.

We can find numerical estimates for the renormalisation (of the
coupling constant) and rescaling (of the energy scale) by comparing
the TCSA and TBA data. Firstly we consider only the energy gaps to
remove the unphysical ground state energy contributions in the TCSA
scheme. Next, since both the overall scale and the value of
the coupling are different, we consider the ratio of two
gaps to find a quantity which is independent of the energy
rescaling and use this to determine the effective coupling
constant. We can then use this in turn to find the energy rescaling. 

\begin{figure}[hbt]
\subfigure[The renormalised coupling constant $\lambda_n g_n(\lambda_n)$
plotted against $\lambda_n$]{\scalebox{0.7}{\includegraphics{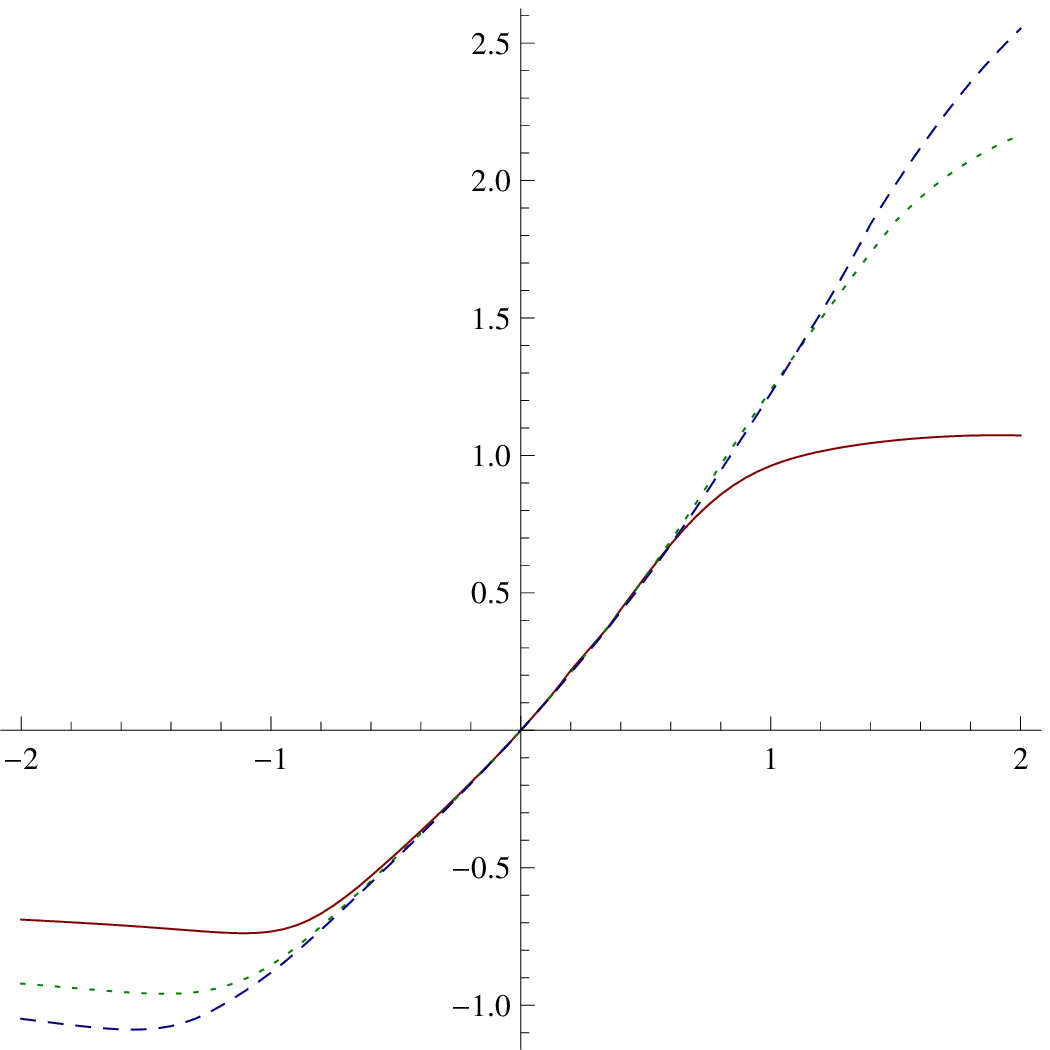}}%
\label{fig:renorm12}}
\hfill
\subfigure[The energy rescaling function $r_n(\lambda_n)$ plotted against
$\lambda_n$]{\scalebox{0.7}{\includegraphics{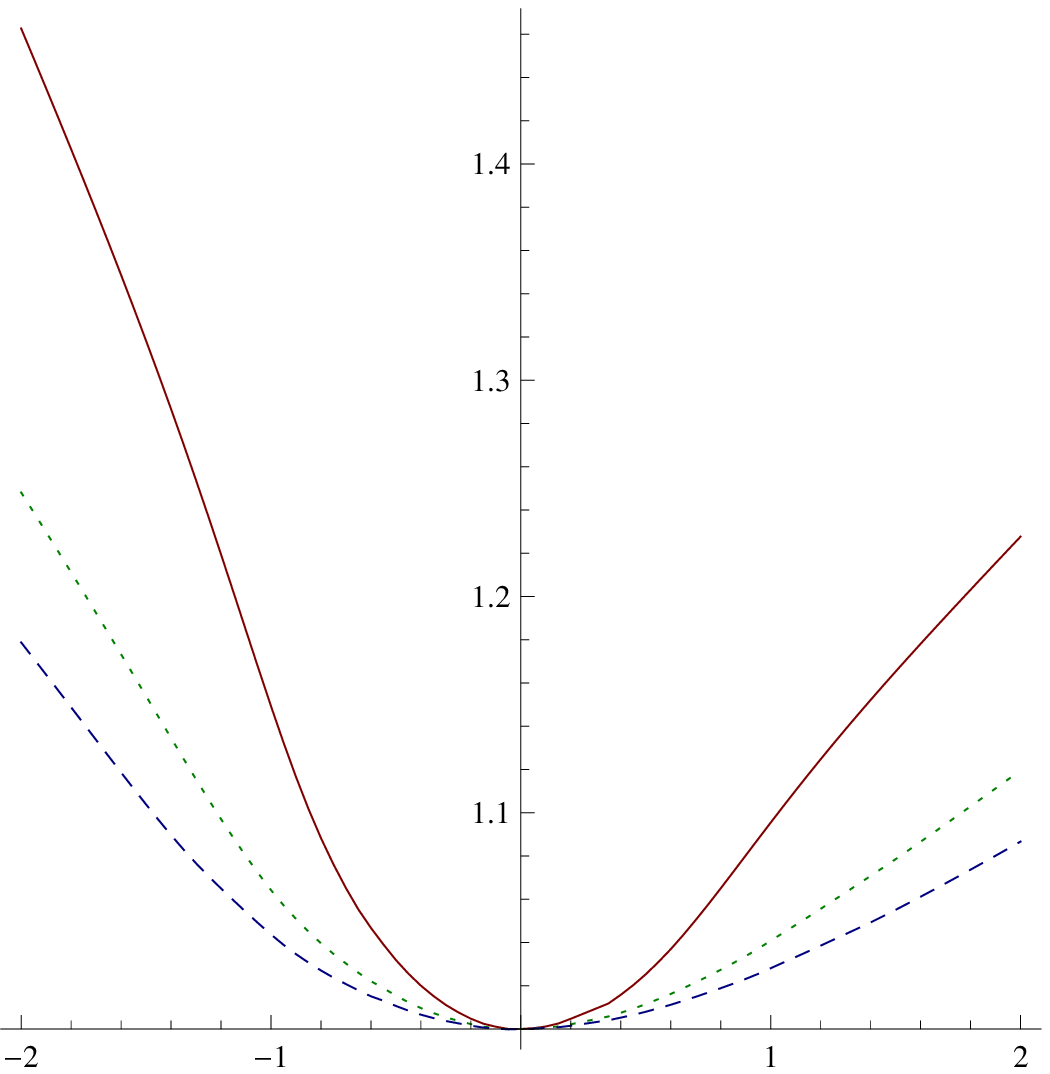}}%
\label{fig:rescaling12}}
\caption{The numerical coupling constant renormalisation and energy
  rescaling found for the tri-critical Ising model at truncation
  levels 6 (red, solid), 14 (green, dotted) and 22 (blue, dashed).} 
\label{figure4}
\end{figure}

If we denote the coupling constant in TCSA by $\lambda_n$ and that in
the TBA by $\lambda_\infty$, and the $n^{th}$ energy gaps by
$\Delta^n_n$ and $\Delta^n_\infty$ respectively, then we can calculate

\be
 f_\infty(\lambda_\infty) = \frac{\Delta^2_\infty}{\Delta^1_\infty}
\;,\;\;
 f_n(\lambda_n) = \frac{\Delta^2_n}{\Delta^1_n}
\;,
\ee
We can then find the effective (TBA) coupling corresponding to a given
TCSA coupling from the function $g_n$ defined by
\be
\lambda_\infty 
= f_\infty^{-1} ( f_n(\lambda_n))
= \lambda_n g_n(\lambda_n)
\;,
\label{eq:gndef}
\ee
and the energy rescaling $r_n(\lambda_n)$ can then be found from
\be
 r_n(\lambda_n) =
 \frac{\Delta^1_n(\lambda_n)}{\Delta^1_\infty(\lambda_\infty)}
\;.
\label{eq:rndef}
\ee
In figure \ref{figure4} we plot these functions for the flows starting
from the $(12)$ boundary condition for three different truncation levels.
It is the aim of this paper to show how these two effects can
predicted from physical and mathematical arguments.

\newpage

\sect{Physical Arguments}

\subsection{The TCSA renormalisation group}

In \cite{FGPTW}, we presented an argument deriving the leading term in
the coupling constant renormalisation based on standard considerations
of the partition function and the operator product expansion of the
perturbing field. We repeat this argument here in a little more
generality, both as a derivation of the coupling constant
renormalisation which we shall check in the next section, but also to
show that this does not in fact give the leading term to the energy
rescaling, as might have been thought was the case.

We consider a strip geometry of width $L$ with complex coordinate $z=x+iy$
with $0\leq y \leq L$. We take 
a general boundary perturbation of the form
\be
\sum_i \mu_i \phi_i(x)
\;,
\label{eq:bp1}
\ee
where $\mu_i$ are the couplings to the boundary fields $\phi_i$ which
have conformal weights $h_i$. We define $y_i = 1 - h_i$. 

We consider a truncation to level $n$ which we can treat using a
projection operator $P_n$ onto states of excitation level $n$ or
below.
We can use the projection operator to restrict the perturbation to the
truncated Hilbert space so that the TCSA approximation to the
perturbation \ref{eq:bp1} 
is then given by 
\be
\sum_i \mu_i (P_n \phi_i P_n)
\;.
\label{eq:bp2}
\ee

The effect of the perturbation is to introduce the following operator
\be
\Pc \exp\left( - \int_{x=-\infty}^\infty \sum_i \mu_i (P_n \phi_i(x) P_n) 
\,\rd x \right)
\;,
\label{eq:pfp1}
\ee
into all expectation values, where $\Pc$ is path ordering.
Mapping the strip to the upper half plane by $w = \exp(\pi z/L)$, this
operator becomes
\be
\Pc \exp\left( - \int_{x=0}^\infty \sum_i \lambda_i (P_n \phi_i(x) P_n) 
\frac{\rd x}{(x\pi)^{y_i}} 
\right)
\;,
\label{eq:pfp1b}
\ee
where $\lambda_i = \mu_i L^{y_i}$ is the dimensionless coupling constant.

Expanding this exponential out to second order we get
\bea
1 &-& \sum_i\lambda_i\int_{x=0}^\infty P_n\phi_i(x)P_n\frac{\rd
  x}{(x\pi)^{y_i}} 
\nonumber\\
 &+& \sum_{jk} \lambda_j\lambda_k \int_{x=0}^\infty
 \int_{x'=0}^x P_n\phi_j(x)P_n\phi_k(x')P_n 
 \frac{\rd x}{(x \pi)^{y_j}}
 \frac{\rd x'}{(x' \pi)^{y_k}}
\;+\; O(\lambda^3)
\;.
\label{eq:2ndo}
\eea
We define the TCSA renormalisation group by the requirement that the
partition function is unchanged when the truncation level is
increased. This introduces a level dependence into the coupling
constants, which we now denote by $\lambda_i(n)$.

Due to translation invariance, we do not need to compare the whole
expression in equation 
\ref{eq:2ndo}, instead we can just compare the integrand with respect
to $x$, at $x=1$, which is
\be
   \sum_i\frac{\lambda_i(n)}{\pi^{y_i}}P_n \phi_i(1) P_n
 - \sum_{jk} \frac{\lambda_j(n)}{\pi^{y_j}}\frac{\lambda_k(n)}{\pi^{y_k}}
   \int_{x'=0}^1 P_n \phi_j(1)P_n \phi_k(x')P_n 
  \frac{\rd x'}{(x')^{y_k}}
+ O(\lambda^3)
\;.
\nonumber\label{eq:pfp2}
\ee
We can further simplify matters by assuming that the Hilbert space
contains a vacuum state $\vac$ and a state $\vec{\phi_i}$
corresponding to the field which satisfy
\be
  \cev{ \phi_i } \phi_j(1) \vac = \delta_{ij}
\;,\;\;
  \cev{\phi_i} \phi_j(1) \phi_k(x) \vac = (1-x)^{x_i-x_j-x_k} C_{ijk}
\;.
\label{eq:phidefs}
\ee
Sandwiching equation
\ref{eq:pfp2} between the states $\cev{\phi_i}$ and $\vac$,
we get 
\be
\lambda_i(n)
- 
\sum_{jk} \lambda_j(n)\lambda_k(n)
\int_{x=0}^1 \cev{ \phi_i } \phi_j(1)\,P_n\,\phi_k(x) \vac
(\pi x)^{y_i-y_j-y_l} \rd x
+ O(\lambda^3)
\;.
\ee
This must be invariant under changes in $n$ so that two second order
in $\lambda$.
Using the fact that
$\cev{\phi_k}\phi_i(1)(P_{n+1}-P_n)\phi_j(x)\vac$ is the coefficient
of $x^{n+1}$ in the expansion of $(1-x)^{h_i-h_j-h_k}$ and is equal
to 
\be
\frac{\Gamma(h_j+h_k-h_i+n+1)}
{\Gamma(h_j+h_k-h_i)\Gamma(n+2)}
= \frac{n^{y_i - y_j - y_k}}{\Gamma(h_j+h_k-h_i)}(1 + O(1/n))
\;,
\ee
we find the TCSA renormalisation group equations
\be
n \frac{\rd \lambda_i}{\rd n}
\simeq
n( \lambda_i(n+1) - \lambda_i(n) )
= \sum_{jk}\lambda_j \lambda_k 
  (n\pi)^{y_i-y_j-y_k} \frac{ C_{ijk}}{\Gamma(h_j+h_k-h_i)}
+ O(\lambda^3)
\;.
\label{eq:tcsade1}
\ee

\subsection{The coupling constant renormalisation}

In the case of the perturbation by a single field with self-coupling
$C$, the TCSA renormalisation group equation (\ref{eq:tcsade1}) becomes 
\be
n \frac{\rd \lambda}{\rd n}
= \frac{ C}{\Gamma(h)(n\pi)^{y} }
  \lambda^2
+ O(\lambda^3)
\;,
\label{eq:diffeq1}
\ee
with solution
\be
  \lambda(\infty) 
= \frac{\lambda(n)}
       {\displaystyle 1 \;\;-\;\; \frac{C}{y\Gamma(h)(n\pi)^y}
        \,\lambda(n)}
\;.
\label{eq:soln1}
\ee
There are two predictions from this calculation - firstly the general
prediction that the coupling renormalisation has a particular scaling
form, namely that the function $g_n$ in equation \ref{eq:gndef} has
the form
\be
g_n(x) = g(x n^{-y})
\;,
\label{eq:gnform}
\ee
and secondly a particular prediction for the 1-loop behaviour,
\be
  g(x) 
= \left(1 - \frac{Cx}{y \Gamma(h) \pi^{-y}}\right)^{-1}
+ O(x^2,n^{-2y})
\label{eq:gn1loop}
\ee
We test these predictions for the tri-critical Ising model in section
\ref{sec:tcimtest1}. 
We will find that the functions $\log( g_n(-x n^y))$ are indeed almost
identical for various values of $n$ and in good agreement with the
prediction, confirming both the prediction of the scaling form and the
approximate numerical expression for this scaling function.

\subsection{The energy rescaling}

We can try to apply the same arguments to derive the energy rescaling,
assuming that this too comes from the operator product of the
perturbing fields giving a contribution to the bare action
(corresponding to the kinetic term in a standard Lagrangian theory).
The Bare Hamiltonian is $(\pi/L)(L_0 - c/24)$. The important part is
the term $L_0$ which is a mode of the energy-momentum tensor
$T(z)$.
This field appears in the operator product of the boundary perturbing
field as
\bea
 \phi(z) \; \phi(w) 
&\sim&
 \frac{1}{(z-w)^{2h}} 
+\frac{C}{(z-w)^h}\phi(w)
+\frac{h/c}{(z-w)^{2h-2}}T(w)
+\ldots
\;,\;\;
\label{eq:ope1}
\\
 T(z) \; \phi(w) 
&\sim&
 \frac{2h}{(z-w)^2}\phi(w)
+\frac{1}{z-w}\phi'(w)
+\ldots
\;,\;\;
\label{eq:ope2}
\eea
where we have also included the OPE of $T$ with the perturbing field.

If we denote the coupling to the perturbing field by $\lambda$ and to
the energy momentum tensor on the boundary by $\lambda_T$ we then 
have $h_T = 2$, $y_T = -1$ and we find
the TCSA renormalisation group equations become
\bea
 n \frac{\rd \lambda}{\rd n} 
&=& 
 \frac{C}{(n\pi)^y\Gamma(h)}\lambda^2 
+(4hn\pi) \lambda\lambda_T 
+\ldots
\;,
\\
 n \frac{\rd \lambda_T}{\rd n} 
&=& 
 \frac{2h/c}{(n\pi)^{1+2y}\Gamma(-2y)} \lambda^2 
+\ldots
\;.
\eea
To find the leading dependence on the induced coupling $\lambda_T$ we
can take $\lambda$ to be constant and using $\lambda_T(\infty)=0$, find
\bea
 \lambda_T(n) 
&=& -\int_{n}^\infty \frac{\rd \lambda_T}{\rd n} \rd n
\nonumber\\
&=& - \frac{2h/c}{\Gamma(-2y)(1+2y)} \frac{\lambda^2}{(n\pi)^{2y+1}}
+ O(\lambda^3)
\;.
\eea
The addition of the term
\be
 \int \mu_T T(x) \rd x
\;,
\ee
to the action will give the following term to the Hamiltonian (after
mapping from the strip to the upper half plane)
\be
  \delta H
= \mu_T T(1)
= \frac{\pi}{L}\left(
 - \frac{2h/c}{\Gamma(-2y)(1+2y)} 
   \frac{\lambda^2}{(n\pi)^{2y}}
   \frac{1}{n}
   \Big(L_0 - c/24\Big) + \hbox{other modes}
  \right)
\;,
\ee
so that the Hamiltonian becomes
\be
H + \delta H
= \frac{\pi}{L}\left( 1
 - \frac{2h/c}{\Gamma(-2y)(1+2y)} 
   \frac{\lambda^2}{(n\pi)^{2y}}
   \frac{1}{n}
 \right)    \Big(L_0 - c/24\Big) + \ldots
\;,
\ee
This gives the rescaling function as
\be
 r_n(\lambda)
= 
1
 - \frac{2h/c}{\Gamma(-2y)(1+2y)} 
   \frac{\lambda^2}{(n\pi)^{2y}}
   \frac{1}{n}
+\ldots
\;.
\ee 
Unfortunately, this has the wrong $n$-dependence - it is one order of
$n$ too small. If we plot $\log(r_n(\lambda) - 1)$ against $\log(n)$
for fixed $\lambda$ for the functions shown in figure 
\ref{figure3}(b), in the tri-critical Ising model, we find that the
prediction for the leading exponent
\be
\log(r_n(\lambda)-1) = \alpha \log(n) + \ldots
\;,
\ee
is approximately $\alpha = -0.85$, which is close to $-2y=-0.8$ and
far from $-1-2y=-1.8$.
As a consequence, we deduce that the coupling of
the perturbing fields to the energy-momentum tensor is not the leading
source of the correction to 
the energy rescaling function. 

Instead we look to the constant ground-state energy contribution
coming from the coupling to the identity operator. As we show in the
next section, this is actually not a constant because of the presence
of the projector $P_n$ and this 
can also give a correction proportional to the bare Hamiltonian.

\subsection{Corrections from the identity operator}
\label{sec:corrid}

In perturbation theory where the perturbing field has a weight greater
than 1/2 one normally ignores the ground state energy as it is a
divergent unphysical quantity. In this case the TCSA cutoff makes the
contribution finite, and further more, the presence of the projector
$P_n$ means it is not constant. 

We are interested in the correction to the coupling to the identity
operator which arises when the expression \ref{eq:pfp2} acts on a
state of excitation level $E$. 
If we denote this state by $\vec E$ and the coupling to the identity
by $\lambda_1(E,n)$, then we are interested in 
the expression 
\bea
\frac{\lambda_1(E,n)}{\pi}
 - \frac{\lambda(n)^2}{\pi^{2y}}
   \int_{x'=0}^1 \cev{E}\phi(1)P_n \phi(x')\vec{E}
  \frac{\rd x'}{(x')^{y}}
\label{eq:pfp3}
\eea
Requiring that this be invariant as we change the truncation level
$n$, we get
\bea
 \frac{\rd \lambda_1(E,n)}{\rd n}
&\simeq&
( \lambda_1(E,n+1) - \lambda_1(E,n) )
\nonumber\\
&=& 
  \frac{\lambda(n)^2}{\pi^{2y}}
   \int_{x=0}^1 \cev{E}\phi(1)(P_{n+1}-P_{n}) \phi(x)\vec{E}
  \frac{\rd x}{(x)^{y}}
\;,
\label{eq:tcsade2}
\eea
where we only take the contribution to the OPE on the right hand side
that comes from the identity channel.
The projector $(P_{n+1}-P_n)$ picks out the state at level $n+1$ in the
action of the field $\phi(x)$ on the state $\vec E$. The action of the
field $\phi(x)$ on the state $\vec E$ has a mode expansion
\be
  \phi(x)\vec E = \sum \phi_m x^{m-h} \vec E
\;,
\ee
so that the term at excitation level $n$ is just the coefficient of
$x^{n+1-E}$. From the operator product expansion (\ref{eq:ope1}), the
coefficient of $x^{n+1-E}$ in $\cev E \phi(1)\phi(x)\vec E$ is
\be
  \frac{\Gamma(n+1-E+2h)}{\Gamma(n+2-E)\Gamma(2h)}
\;,
\ee
This means that the leading contribution to the RG equation
(\ref{eq:tcsade2} for the coupling $\lambda_1$ is 
\bea
  \frac{\rd \lambda_1(E,n)}{\rd n}
&\simeq&
  \frac{\lambda(n)^2}{\pi^{2y}}
   \int_{x=0}^1 
  \frac{\Gamma(n+1-E+2h)}{\Gamma(n+2-E)\Gamma(2h)}
  x^{n+1-E}
  \frac{\rd x}{(x)^{y}}
\nonumber\\
&\simeq&
 \frac{\lambda^2}{\pi^{2y}}
 \frac{(n-E)^{2h-2}}{\Gamma(2h)}
+ \ldots
\nonumber\\
&\simeq&
 \frac{\lambda^2}{(n\pi)^{2y}}
 \frac{1}{\Gamma(2h)}
 \left(
1 \;+\; 2y\frac{E}{n}
+ \ldots
 \right)
\label{eq:2ndo3}
\eea
The first term is the constant contribution which we shall now ignore;
the second is the excitation-level dependant term which we are
interested in. 
We can integrate equation (\ref{eq:2ndo3}) using the boundary condition
$\lambda_1(E,\infty)=0$, to find
\bea
  \lambda_1(E,n)\big|_E
&=&
  -\int_{n}^\infty 
 \frac{\lambda^2}{\pi^{2y}}
 \frac{1}{\Gamma(2h)}
2yE n^{-2y-1}
\rd n
\nonumber \\
&=& \frac{\lambda^2}{(n\pi)^{2y}}
    \frac{1}{\Gamma(2h)} E
\;.
\label{eq:2ndo4}
\eea
We can now replace the excitation level $E$ by the operator $(L_0-c/24)$, as
$E$ is approximately the eigenvalue of $(L_0 - c/24)$, and we finally
find we have generated a term proportional to the bare Hamiltonian.
We have two equivalent ways of understanding this term which lead to
the same expression for the energy rescaling $r_n(\lambda)$

Firstly, we can see that this coupling will lead directly to a change
in the Hamiltonian; the new Hamiltonian is
\be
  \frac{\pi}{L} \left[
  (1 + \frac{\lambda^2}{(n\pi)^{2y}} \frac{1}{\Gamma(2h)} )
  (L_0 - c/24)
  + \frac{\lambda}{\pi^y}\phi(1)
  \right]
\;,
\ee
giving the energy rescaling function as
\be
  r_n(\lambda) =  1 + \frac{\lambda^2}{(n\pi)^{2y}}
  \frac{1}{\Gamma(2h)}  + \ldots
\;,\;\;
  r(x) =  1 + \frac{x^2}{\pi^{2y}}
  \frac{1}{\Gamma(2h)}  + \ldots
\;.
\label{eq:rn}
\ee
This has the correct $n$ dependence (in agreement with the numerical
data) and is also a very good fit to the actual rescaling
function as we see in section \ref{sec:tcimtest2}. 

An alternative viewpoint of the energy rescaling function is that it
represent an effective change in the geometry of the system. 
Coordinate transformations are implemented in CFT by changes to the
action (see e.g. \cite{CardyLH1}). The coordinate change $x^\mu \to
\alpha^\mu + \alpha^\mu$ corresponds to the change in the action
\be
\delta S = - \frac{1}{2\pi}\int T_{\mu\nu}\partial^\mu \alpha^\nu
\;\rd^2 x
\;.\ee
The energy-dependant correction to the identity operator we have just
calculated can also be put in this form.
On the upper-half-plane, with complex coordinate $w=r\exp(i\theta)$,
the correction is  
\be
\delta S
=
- \frac{1}{\pi}\int \lambda_1(E,n) \frac{\rd r}{r}
=
- \frac{\lambda^2}{(n\pi)^{2y}\Gamma(2h)} \int L_0 \frac{\rd r}{r} \;.
\label{eq:corr1}
\ee
Firstly, we note that on the upper half plane
\be
  L_0 = \frac{\pi}{L} \int_{\theta=0}^\pi (w^2 T(w) + \bar w^2\bar T(\bar
  w))\frac{\rd\theta}{2\pi}
\;.
\label{eq:l0}
\ee
Combining equations (\ref{eq:corr1}) and (\ref{eq:l0}) and
transforming to the strip with coordinate $z=x+iy$, the correction to
the action becomes 
\be
\delta S
= - \frac{\lambda^2}{(n\pi)^{2y}\Gamma(2h)}\int T_{xx}\frac{\rd x\,\rd
  y}{2\pi}
\;,
\label{eq:ds2}
\ee
which in turn corresponds to the change in coordinates
\be
\pmatrix{x \cr y}
\to
\pmatrix{x \left( 1 + \frac{\lambda^2}{(n\pi)^{2y}\Gamma(2h)}\right)
  \cr y}
= 
\pmatrix{r_n(\lambda) x 
  \cr y}
\;.
\ee
If we consider the theory on a strip of length $R$ then this change in
coordinates is an effective increase in the length by a factor
$r_n(\lambda)$. To the order in $\lambda$ to which we are working,
this is equivalent to a reduction in the strip width by the same
factor and consequently a rescaling of the eigenvalues of the
Hamiltonian by the same factor, $r_n(\lambda)$, exactly in accordance
with equation (\ref{eq:rn}).

\sect{Mathematical Arguments}

It is also possible to derive the coupling constant renormalisation
and energy rescaling by examining the lowest three eigenvalues of the
perturbed, truncated Hamiltonian directly. This gives exact
expressions in terms of integrated four-point functions of the
conformal field theory. We can then find the large-$n$ behaviour of
these eigenvalues by a saddle-point approximation. It turns out that
these are given in terms of the crossed four-point functions and the
leading large-$n$ behaviour is given by the leading terms in the
crossed four-point functions, which are in turn given by the operator
product expansion coefficients. In this way we see that the
``physical'' arguments of the previous section are indeed correct.
Furthermore, we can show that the corrections we have found are the
leading order corrections for all the eigenvalues of the perturbed
Hamiltonian, not just the lowest three.

We start from the expression (\ref{eq:Hpcft3}) for the dimensionless TCSA
Hamiltonian:
\newcommand{\tl}{{\tilde\lambda}}
\be
  \Hc_n = (L_0 - \frac c{24}) + \tl P_n\phi(1)P_n
\;.
\ee
We have introduced $\tilde\lambda = \lambda\pi^{-y}$ for notational
convenience. 

For simplicity we consider a case where the Hilbert space is a single
highest weight representation of the Virasoro algebra with conformal
weight $H$, and that the two lowest level states are
\be
 \vec{\psi}
\;,\;\;
 L_{-1} \vec{\psi}
\;.
\ee
We will assume that the the energy-rescaling and
coupling constant renormalisation take the forms (\ref{eq:gnform}) and 
\be
g(x) = 1 + b \pi^{-y} x + O(x^2,n^{-2y})
\;,\;\;
r(x) = 1 + a \pi^{-2y} x^2 + O(x^3,n^{-3y})
\;.
\label{eq:gr}
\ee
We also assume that the first energy gap takes the exact form
\be
  \Delta_1(\lambda) 
= E_1(\lambda)-E_0(\lambda)
= 1 + \alpha\tl + \beta\tl^2 + O(\lambda^3)
\;,
\label{eq:delta1}
\ee
so that the TCSA approximation at truncation level $n$ is given by
\bea
\Delta_1^{n}(\lambda)
&=& r(\lambda n^{-y})\Delta_1(\lambda g(\lambda n^{-y}))
\nonumber\\
&=& 1 + \alpha\tl
+ \tl^2( \beta + a n^{-2y} + b \alpha n^{-y})
+ O(\lambda^3,n^{-3y})
\;.
\label{eq:ndep}
\eea
This means that we can find the coefficients $a$ and $b$ appearing in
the rescaling and renormalisation  functions (\ref{eq:gr}) from the
$n$ dependence in the second order term in the first energy gap. 
From the physical arguments of the previous section, we expect
\be
 a = \frac{1}{\Gamma(2h)}
\;,\;\;
 b = \frac{C}{y\Gamma(h)}
\;.
\label{eq:ab}
\ee

By standard perturbation theory, we find the following expressions for
the ground state and first excited state energies, to second order:
\bea
  E_0
&=& 
H + \tl C'
- (\tl C')^2 \int_0^1(F^0 - \frac{1}{z^h})\frac{\rd
  z}{z^y}
\;,\;\;
\nonumber\\
  E_1
&=&
(H+1) + \tl C'\left(1 + \frac{(h-1)h}{2H}\right)
\nonumber
\\
&&+ (\tl C')^2\left[
  \frac{h^2}{2H} - 
\int_0^1 \frac{\rd z}{z^y}
\left( F^1 - \frac{h^2/(2H)}{z^{2-y}}
      - \frac{(2H+h(h-1))^2}{(2H)^2z^{1-y}}
 \right)
\right]
\;.
\label{eq:ints1}
\eea
Here $C'$ is the boundary coupling constant in the operator product
expansion
\be
  \phi(x) \vec\psi = \frac{C'}{ x^{h}} \vec\psi + \ldots
\;,
\ee
and the function $F^0$ is the four point chiral block 
\be
  F^0(z) 
= 
\raisebox{-2.5mm}{\mbox{
$\cbb{2300}{\psi}{\phi}{\psi}{\phi}{\psi}{1}{z}
\;\;
$}}
= z^{-h}(1 + \ldots)
\;,
\ee
that appears in the boundary two-point function on the strip.
Since we have chosen boundary conditions for which the Hilbert space
of the model on the strip has only a single 
representation with highest weight $\vec\psi$, the boundary two point
function is given as the product of the structure constants and the
single chiral block with the representation $\psi$ in the intermediate
channel: 
\be
\langle \phi(1) \phi(z) \rangle_{\mathrm{strip}}
= \cev\psi \phi(1) \phi(z) \vec\psi
= (C')^2 F^0(z)
\;.
\ee
$F^1$ is the function that appears in the two point function of the
field $\phi(z)$ in the first excited state,
\bea
&&
  \frac{ \cev\psi L_1 \phi(1) \phi(z) L_{-1} \vec\psi}{\cev\psi
  L_1 L_{-1} \vec\psi} 
= (C')^2 F^1(z)
\;,\;\;
\nonumber\\
F^1(z) &=& \frac{1}{2H}\left\{
   \left[ z(z-1)\frac{\rd}{\rd z} + 2zh-1\right]
   \left[ (z-1)\frac{\rd}{\rd z} + 2h\right]  
   + 2 H\right\} F^0
\;.
\eea
If $h$, the conformal dimension of the perturbing field, is greater
than or equal to 1/2 then the integrals in the expressions 
(\ref{eq:ints1}) diverge, but their difference, the energy gap
$\Delta_1$, is finite.
We can read off the coefficients $\alpha$ and $\beta$ that appear in
(\ref{eq:delta1}) as
\bea
  \alpha
&=& \frac{h(h-1)}{2H} C'
\;,\;\;
\label{eq:alpha}\\
  \beta
&=&
  (C')^2\left\{
 \frac{h^2}{2H}
- \int_0^1 \frac{dz}{z^y}
\left[
 \left(F^1 - \frac{h^2/(2H)}{z^{2-y}}
      - \frac{(2H+h(h-1))^2}{(2H)^2z^{1-y}}
 \right)
 - \left( F^0 - \frac{1}{z^h}
 \right)
\right]
\right\}
\nonumber\\
\label{eq:beta}
\eea
These are the exact coefficients in the perturbative expansions of the
energy gap; we are interested in the $n$-dependence in the TCSA
approximation (\ref{eq:ndep}). The $n$-dependence comes from replacing
the functions $F^0$ and $F^1$ in (\ref{eq:beta}) by the TCSA
truncations. We see that $\alpha$ is independent of $n$ but $\beta$
does depend on $n$; we shall denote the TCSA truncation by $\beta(n)$
and we expect from (\ref{eq:ndep}) that
\be
\beta(n)
= 
\beta + a n^{-2y} + b \alpha n^{-y}
+ O(n^{-3y})
\;.
\ee
As before, in section \ref{sec:corrid}, the TCSA truncations of the
functions $F^0$ and $F^1$ come from restricting their Taylor expansions
to include only modes up to power $z^{n-h}$. It is easier to analyse the
change in the functions as the TCSA level is increased than to find
the $n$-dependence directly, so we consider the change in $\beta$
given by the taking just the coefficient of $z^{n+y}$ in the integrand
of (\ref{eq:beta})
\bea
  \frac{\rd \beta}{\rd n}
&\simeq&
  \beta(n+1) - \beta(n)
=
  \frac{(C')^2}{n}
\oint \frac{ \rd z}{2\pi i z^{n+1}}
 \left[
  z^h (F^0 - F^1)
\right]
\;.
\label{eq:df1}
\eea
The contour in (\ref{eq:df1}) is a small circle around the
origin. The function $z^h(F^1-F^0)$ is single valued around the origin
but has a cut from $z=1$ to $z=\infty$, 
and can be expanded in
increasing powers of $(1-z)^{-1}$.
We use the result that
\be
\oint \frac{\rd z}{2\pi i z^{n+1}} (1-z)^{-\alpha}
= \frac{ \Gamma(n+\alpha) }{ \Gamma(\alpha) \Gamma(n+1) }
= \frac{n^{\alpha-1}}{\Gamma(\alpha)} ( 1 \;+\; O(1/n) \; )
\;,
\ee
to see that the dependence of the function (\ref{eq:df1}) on $n$ is
determined by the expansion of the integrand in powers of $(1-z)$.
The expansion of $F^0$ in $(1-z)$ is determined by taking the
alternative conformal block expansion of the boundary two point function:
\bea
 (C')^2  F^0(z) 
&=& 
\raisebox{-2.5mm}{\mbox{\cbI{2500}{\psi}{\phi}{\phi}{\psi}{\id}st  }}
\; + \;
C C' 
\raisebox{-2.5mm}{\mbox{\cbI{2500}{\psi}{\phi}{\phi}{\psi}{\phi}st }}
\; + \; \ldots
\nonumber
\\[3mm]
&=&
  (1-z)^{-2h}\left\{1 + O(1-z)^2\right\}
\;+ \;
  CC' (1-z)^{-h}\left\{1 + O(1-z) \right\}
\;+ \; \ldots
\nonumber
\\
\label{eq:crossed}
\eea
The further terms in (\ref{eq:crossed}) correspond to further fields
in the operator product expansion of $\phi(x)$ with itself. We assume
that these are less singular than the two shown, as is the case for
perturbations by the unitary minimal model field $\phi_{13}$.
Calculating the integral (\ref{eq:df1}) we get
\be
\frac{\rd\beta}{\rd n}
=
 \frac{2(h-1)}{\Gamma(2h)}n^{-2y-1}
+
 C C' \frac{h(h-1)}{2H\Gamma(h)}n^{-y-1}
+
 \ldots
\;,
\label{eq:df2}
\ee
where it is worth noting that the leading term which one would expect
to have dependence $n^{2h-2}=n^{-2y}$ has zero
coefficient and the term in $n^{-2y-1}$ is the sub-leading term.
We can now integrate (\ref{eq:df2}) to find
\bea
\beta(n)
&=& \beta(\infty) - \int_n^\infty \frac{\rd\beta}{\rd n}\,\rd n
\nonumber\\
&=& 
 \beta(\infty)
+
 \frac{1}{\Gamma(2h)}n^{-2y}
+
 C C' \frac{h}{2Hy\Gamma(h)}n^{-y}
+
 \ldots
\;.
\label{eq:betan}
\eea
This exactly agrees with the expected form (\ref{eq:ndep}), so that
the rigorous mathematical derivation of the leading $n$-dependence of
the TCSA approximation to the first energy gap agrees with the
heuristic derivation of the coupling constant renormalisation and
energy rescaling from physical arguments.

It is important to note that these effects are independent of the
representation $\psi$ chosen, and so, in particular, they should apply
to all of the flows considered by Feverati et al.

It is in fact possible, by considering the effect of replacing the
state $L_{-1}\vec\psi$ by suitable states at arbitrary levels, to
extend this calculation to cover all energy gaps, not just the first
energy gap. This means that if $\Delta_i^n(\lambda)$ is the TCSA
approximation to the $i$-th energy gap, then one can prove that \be
\Delta_i^n(\lambda) = r(\lambda n^{-y})\Delta_i(\lambda g(\lambda
n^{-y})) + O(\lambda^3) \;, \ee where the functions $r$ and $g$ are
those given in (\ref{eq:gr}) with coefficients (\ref{eq:ab}).

\sect{Checks in the tri-critical Ising model}
\label{sec:tcimcheck}

\begin{figure}
\begin{center}
\input{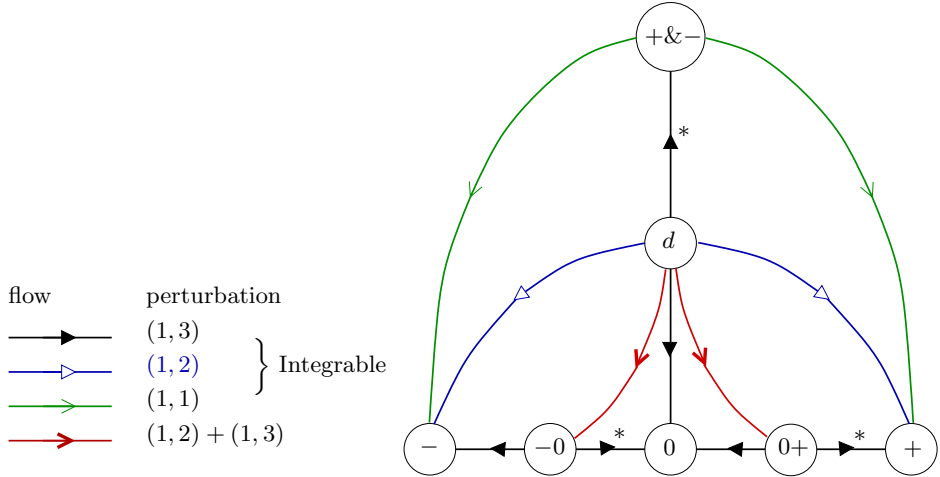}
\end{center}
\caption{The space of boundary flows in the tricritical Ising model}
\label{figure1}
\end{figure}

As mentioned before, we shall perform most of our checks in the
tri-critical Ising model because the excited state spectra of the
strip with a perturbed boundary condition has been studied in detail
using the TBA approach by Feverati et al. \cite{FPR1,FPR2,Feverati1}
allowing us to compare the TCSA results with the `exact' TBA
spectrum.
The tri-critical Ising model is a unitary conformal field theory with
central charge 
$7/10$ and the Virasoro algebra has six unitary highest weight
representations listed in table \ref{table1}. 
The model has six
fundamental, or ``Cardy'', conformally invariant boundary conditions
\cite{Cardy89,Chim95} 
listed in table \ref{table2}.
These can be labelled
either by representation of the Virasoro algebra or by the allowed
values of the boundary spins in its realisation as a spin-1 Ising model.
The boundary flows were first given by Affleck in \cite{Affleck1} and
are shown in figure \ref{figure1}.

\begin{table}[bht]
\renewcommand{\arraystretch}{1.4}
\[
\begin{array}{c|c|c|c|c|c|c}
\hbox{Virasoro label} & (11) & (21) & (31) & (12) & (13) & (22) 
\\[1mm]
\hline
\hbox{Conformal weight} & 0 & \frac 7{16} & \frac 32 & \frac1{10}
& \frac{3}{5} & \frac 3{80} 
\\[1mm]
\hline
\hbox{Boundary spins} & (-) & (0) & (+) & (-0) & (0+) & (-0+)=(d) 
\\[1mm]
\hline
\hbox{Boundary fields} & (11) & (11),(31) & (11) &
(11), {(13)} 
& 
(11), {(13)} 
& 
(11), {(13)},{(12)},(31) 
\end{array}\]
\caption{The representations of the Virasoro algebra in
  the tricritical Ising model and properties of the corresponding
  conformal boundary conditions} 
\label{table1}
\label{table2}
\end{table}

From these boundary conditions and their flows one can easily construct
superpositions of boundary conditions and further flows using the
action of topological defects \cite{GrahamWatts}. 
In the case in hand, the topological defects and the boundary
conditions are labelled by representations of the Virasoro algebra,
and both the action of the defects on the boundary conditions and the
representation content of the strip Hilbert space are given in terms
of the fusion of representations. If we denote the representations of
the Virasoro algebra by $a$, $b$ etc, the defects by $D_a$ and the
boundary conditions by $B_a$ then the defects act as
\be
  D_a \cdot B_b = B_{a * b} = \sum_c N_{ab}{}^c B_c
\;,
\ee
where $N_{ab}{}^c$ are the Verlinde fusion numbers. Likewise, the
representation content of a strip with conformal boundary conditions
$a$ and $b$, which we denote $\cH_{(a,b)}$ is
\be
\cH_{(a,b)}
= \oplus_c N_{ab}{}^c L_c
\;,
\ee
where $L_a$ is the highest weight representation with label $a$.
The commutativity of the fusion algebra means that the strip with
boundary conditions $(a,b*c)$ has the same Hilbert space as the strip
with boundary conditions $(b*a,c)$, which we can interpret as the
statement that the action of a defect of type $b$ on either of the two
boundary conditions $a$ or $c$ leaves the Hilbert space unchanged.
This means that 
the six separate flows considered by Feverati et al. can now be
group into three pairs which can be found as the spectrum of the strip
with the basic set of flows (\ref{eq:basicflow}) on one edge and one
of the three `fixed' conformal boundary conditions $(r1)$ on the
other. By considering the second boundary conditions as the action of
the defect $D_{(r1)}$ on the $(11)$ boundary condition, they have
alternative interpretations as the spectra of strips with perturbed
boundary conditions of type $(12)$, $(13)$ and $(22)$ coupled with an
undeformed $(11)$ boundary condition on the second edge as we show in
table \ref{tab:equiv}.

\begin{table}[htb]
\[
\begin{array}{l|l|l}
\hbox{Spectral flow}
&
\hbox{Strip configuration 1}
& 
\hbox{Strip configuration 2}
\\
\hline
(11)\leftarrow(12)\to(21)
&
(11;11) \leftarrow (12;11) \to (21;11)
&
(11;11) \leftarrow (12;11) \to (21;11)
\\
(31)\leftarrow(13)\to(21)
&
(31;11) \leftarrow (13;11) \to (21;11)
&
(11;31) \leftarrow (12;31) \to (21;31)
\\
(21)\leftarrow(22)\to(11){+}(31)
\!\!&
(21;11)\leftarrow(22;11)\to(11{+}31;11)
\!\!&
(11;21) \leftarrow (12;21) \to (21;21)
\end{array}
\]
\caption{The possible interpretations of the flows
considered by Feverati et al. 
Configuration one has the boundary condition $(11)$ on one side of the
strip; configuration two has a fixed boundary condition of type $(r1)$
on one side and the basic flow $(11)\leftarrow(12)\to(21)$ on the other.}
\label{tab:flows}
\label{tab:equiv}
\end{table}

\subsection{The coupling constant renormalisation}

\label{sec:tcimtest1}

\begin{figure}[hbt]
\subfigure[The scaling functions $\log g_n(-x n^y)$ and the prediction for
$\log(g(-x))$ plotted vs. $\log(-x)$ i.e. for negative coupling
constant]{\scalebox{0.7}{\includegraphics{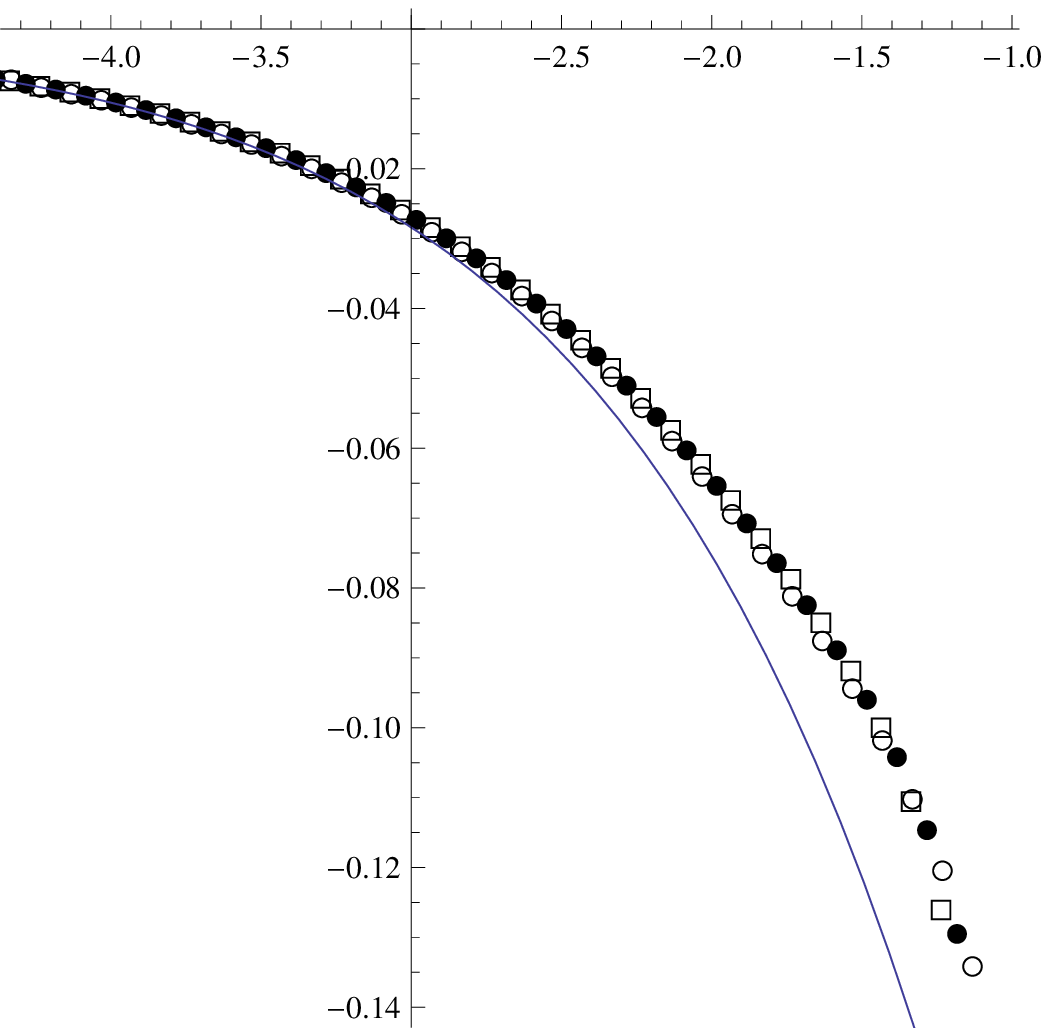}%
\label{fig:scaling12}
\label{fig:gn}}
}%
\hfill
\subfigure[The scaling functions $\log g_n(x n^y)$ and the prediction for
$\log(g(x))$ plotted vs. $\log(x)$ for positive coupling constant]{%
\scalebox{0.7}{\includegraphics{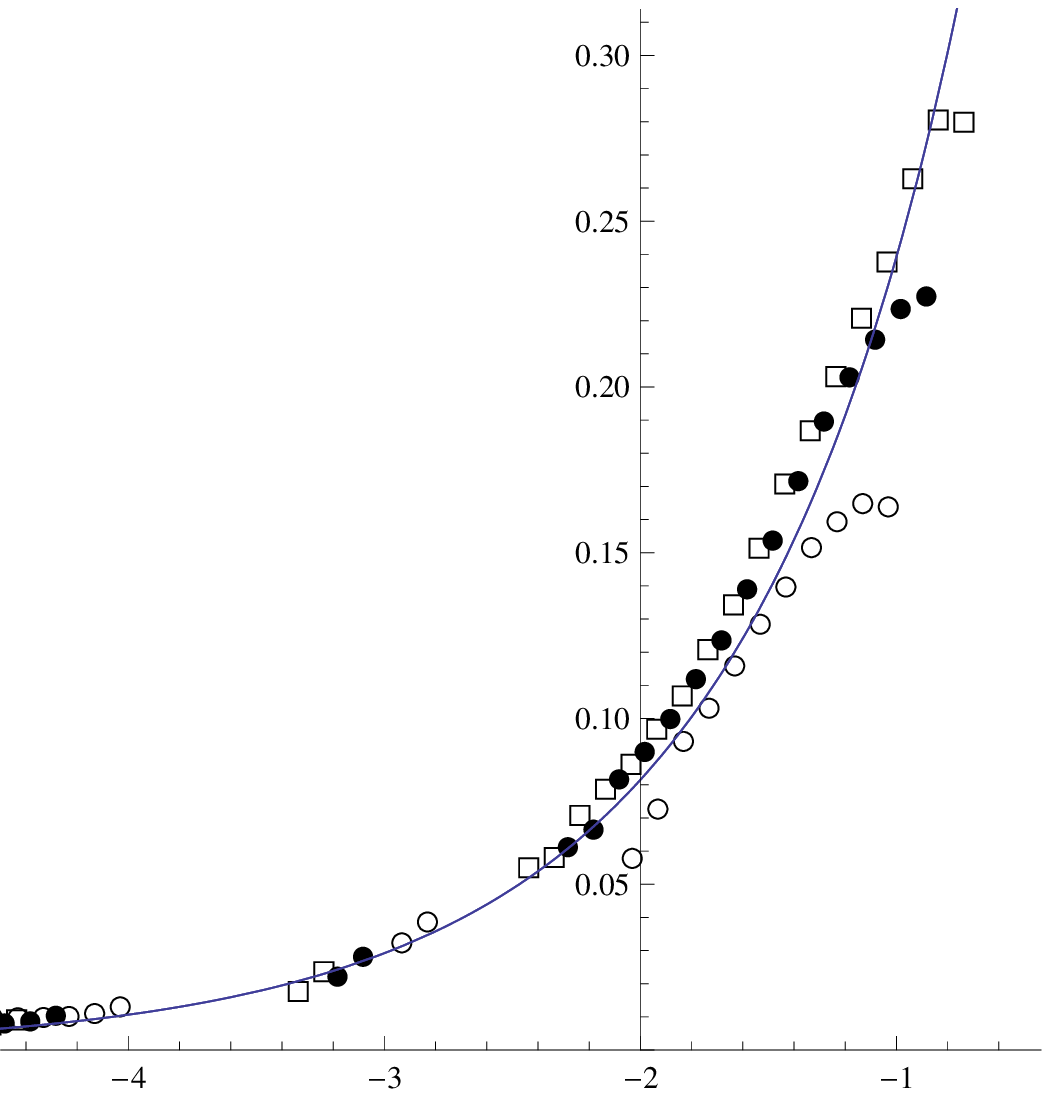}%
\label{fig:scaling12p}}
}
\caption{The numerical coupling constant renormalisation found for the
 tri-critical Ising model with $(11)$ boundary condition on the other
 edge, for truncation levels 8
 ($\circ$) 15 ($\bullet$) and 22 ($\square$) together with the 1-loop
 predictions (\ref{eq:gn1loop}) shown as a solid line. }
\label{fig:gnplot}
\end{figure}

In the tri-critical Ising model we first consider the 
renormalisation of the coupling $\lambda$ of the perturbing field
$\phi_{3/5}$ which generates the flows
\be
 (11) 
\stackrel{-\lambda\phi}{\longleftarrow}
(12)
\stackrel{+\lambda\phi}{\longrightarrow}
(21)
\;.
\label{eq:flows}
\ee
We recall that there are two predictions from this calculation -
firstly the general prediction that the coupling renormalisation has a
particular scaling form, namely that the function $g_n$ in equation
\ref{eq:gndef} has 
the form
\be
g_n(x) = g(x n^{-y})
\ee
and secondly a particular prediction for the 1-loop behaviour,
\be
  g(x) 
= \left(1 - \frac{Cx}{y \Gamma(h) \pi^{-y}}\right)^{-1}
+ O(x^2,n^{-2y})
\label{eq:gn1loop2}
\ee
The values for the tri-critical Ising model are
\be
C 
= -\frac{\Gamma(-\tfrac 35) \Gamma(\tfrac 25 )^{1/2}}
       {\Gamma(\tfrac 15)\Gamma(-\tfrac 65)^{1/2}}
= 0.544542...
\;,\;\;
h = 3/5
\;,\;\;
y = 2/5
\;.
\ee
For the TCSA approach, we need to specify two boundary conditions. 
On one we of course take the perturbed boundary conditions (\ref{eq:flows});
on the other we can take any of the ``fixed'' boundary conditions
$(r1)$ and we will obtain one of the
cases investigated by Feverati et al. 
For simplicity, we start with the boundary condition $(11)$ on the
other.
For the flow with negative $\lambda$, that is for $(11)\leftarrow(12)$,
we plot the functions
$\log( g_n(-x n^y))$ against $\log(-x)$ for several values of $n$
together with the 1-loop prediction for $\log(g(-x))$ in equation
(\ref{eq:gn1loop2}). These are shown in figure \ref{fig:gn}
We see that the functions $\log( g_n(-x n^y))$ are indeed almost identical
for various values of $n$ and in good agreement with the prediction,
confirming both the prediction of the scaling form and the approximate
numerical expression for this scaling function.
For positive values of $\lambda$, 
we plot 
$\log( g_n(x n^y))$ against $\log(x)$ for the same values of $n$,
again 
together with the 1-loop prediction for $\log(g(x))$ in equation
(\ref{eq:gn1loop2}) in figure \ref{fig:scaling12p} with the same results
for small values of $\lambda$.

The scaling form does clearly break down for larger values of
$\lambda$ positive, where the functions $g_n(x n^y)$ have
$n$-dependent maxima at $n$-dependent values of $x$. This corresponds
physically to the boundary condition approaching close to the $(21)$
fixed point but then moving away in the direction of the $(13)$ fixed
point. Presumably the flow in this region is governed by the critical
exponents around the $(21)$ fixed point and different arguments are
required to analyse this behaviour, but we can make some suggestions
about the energy-level-dependence of the position of the fixed point,
which we do later in this section.

\begin{figure}[hbt]
\subfigure[The TCSA coupling constant renormalisation
 $\log(g_n(-xn^y))$ plotted against $\log(-x)$ for the
 tri-critical Ising model and choices
 of second boundary condition 
$(11)=(-)$, ($\circ$, level 22), 
$(21)=(0)$ ($\bullet$, level 14)
and $(31)=(+)$ ($\square$, level 18) together
with the 1-loop  prediction (\ref{eq:gn1loop}) shown as a solid
line.]{\scalebox{0.7}{\includegraphics{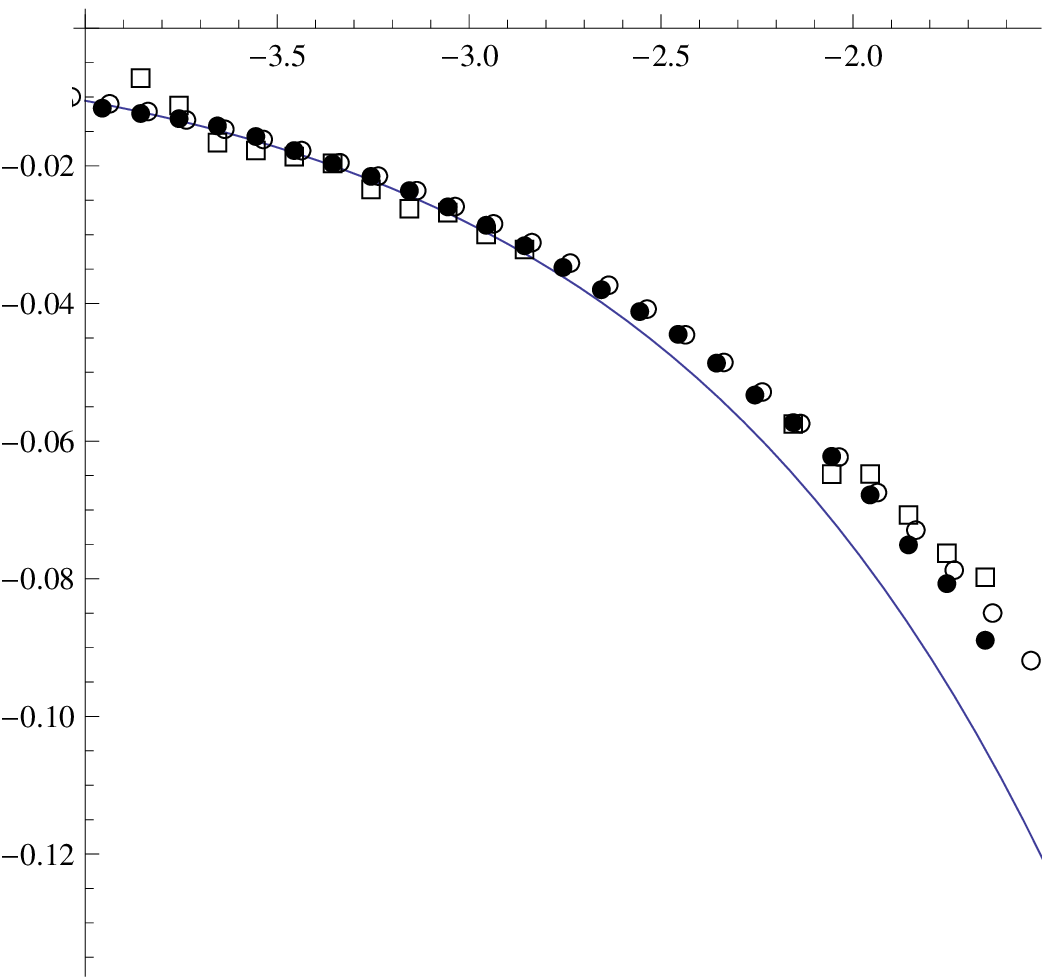}%
\label{fig:scaling112131}}
}%
\hfill
\subfigure[The energy rescaling function $\log r_n(-x n^y)$ and the
  prediction for $\log(r(-x))$
  plotted vs. $\log(-x)$ for negative coupling for the flow
  $(11;11)\leftarrow(12;11)$ 
for truncation levels 8
 ($\circ$) 15 ($\bullet$) and 22 ($\square$).
]{\scalebox{0.7}{\includegraphics{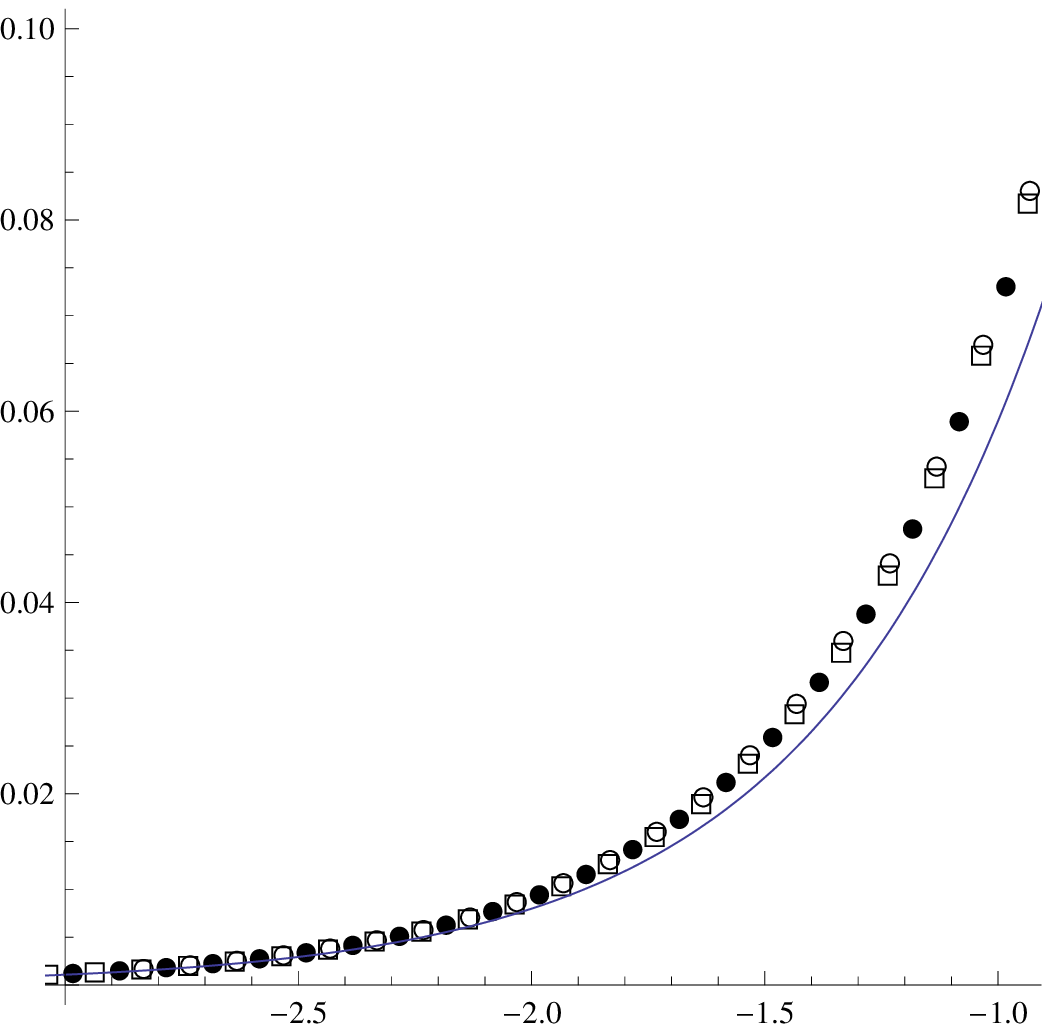}%
\label{fig:rescale12}}
}%
\caption{The coupling constant and energy rescaling functions} 
\label{fig:112131}
\end{figure}

Finally we can check whether the coupling constant renormalisation and
energy re-scalings depend on the second boundary condition on the
strip. In figure \ref{fig:scaling112131} we present the coupling constant
renormalisation as calculated from the three different choices of
second boundary condition, $(11)$, $(21)$ and $(31)$. We see that,
modulo numerical inaccuracies, the three different TCSA strip
configurations give the same coupling constant renormalisation, as
would be expected on physical grounds, and that this is in agreement
wit the 1-loop calculation.

\subsection{The energy rescaling}
\label{sec:tcimtest2}

We can also check the prediction \ref{eq:rn} for the form of the
energy rescaling in the tri-critical Ising model.
This is shown in figure \ref{fig:rescale12} where $\log(r(-x))$ is
plotted together with the numerical estimates for $\log(r_n(-xn^y))$
for various different truncation levels and for different choices of
the non-perturbed boundary condition. This confirms both the scaling
form of $r_n$ and also the numerical coefficient calculated in this
section.

\sect{Checks in the Yang-Lee model}

\begin{figure}[hbt]
\begin{center}
\vspace{3mm}
\begin{tabular}{cc}
\epsfysize=7cm 
\epsfbox{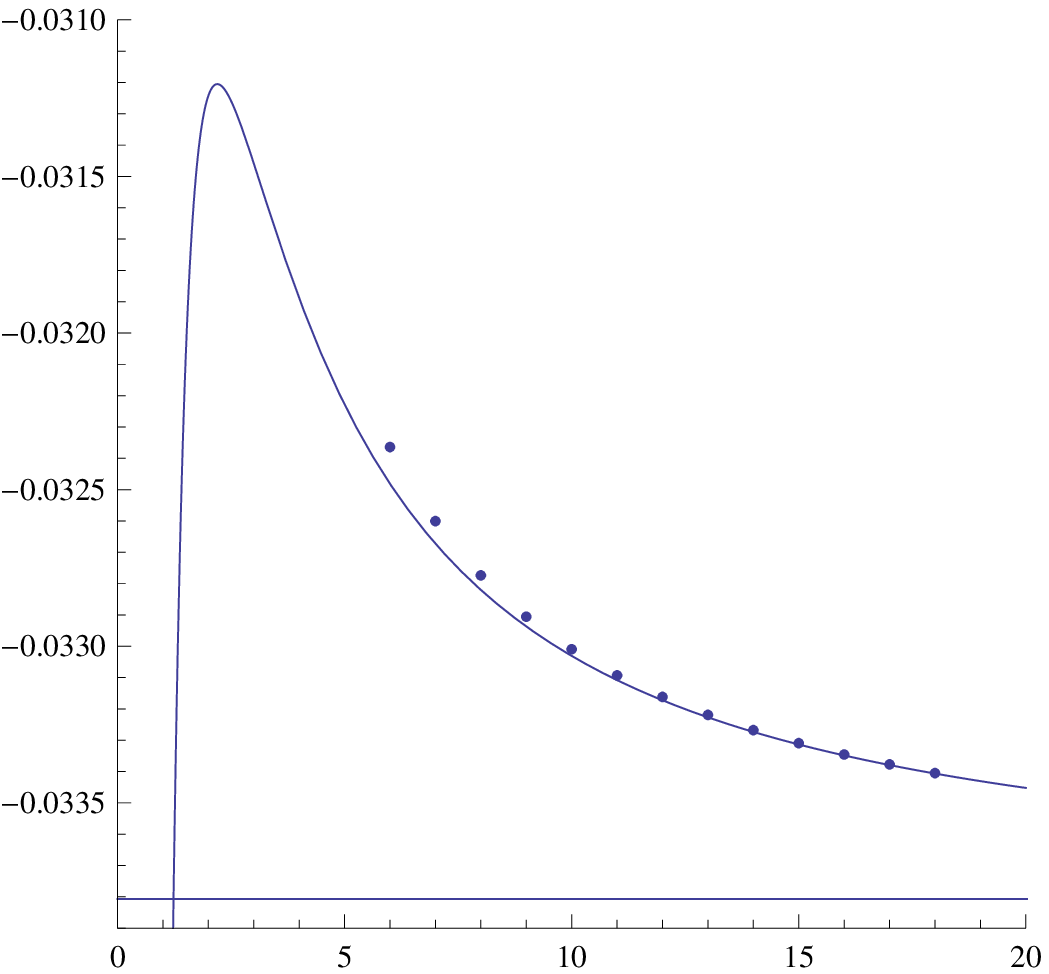}
&
\epsfysize=7cm 
\epsfbox{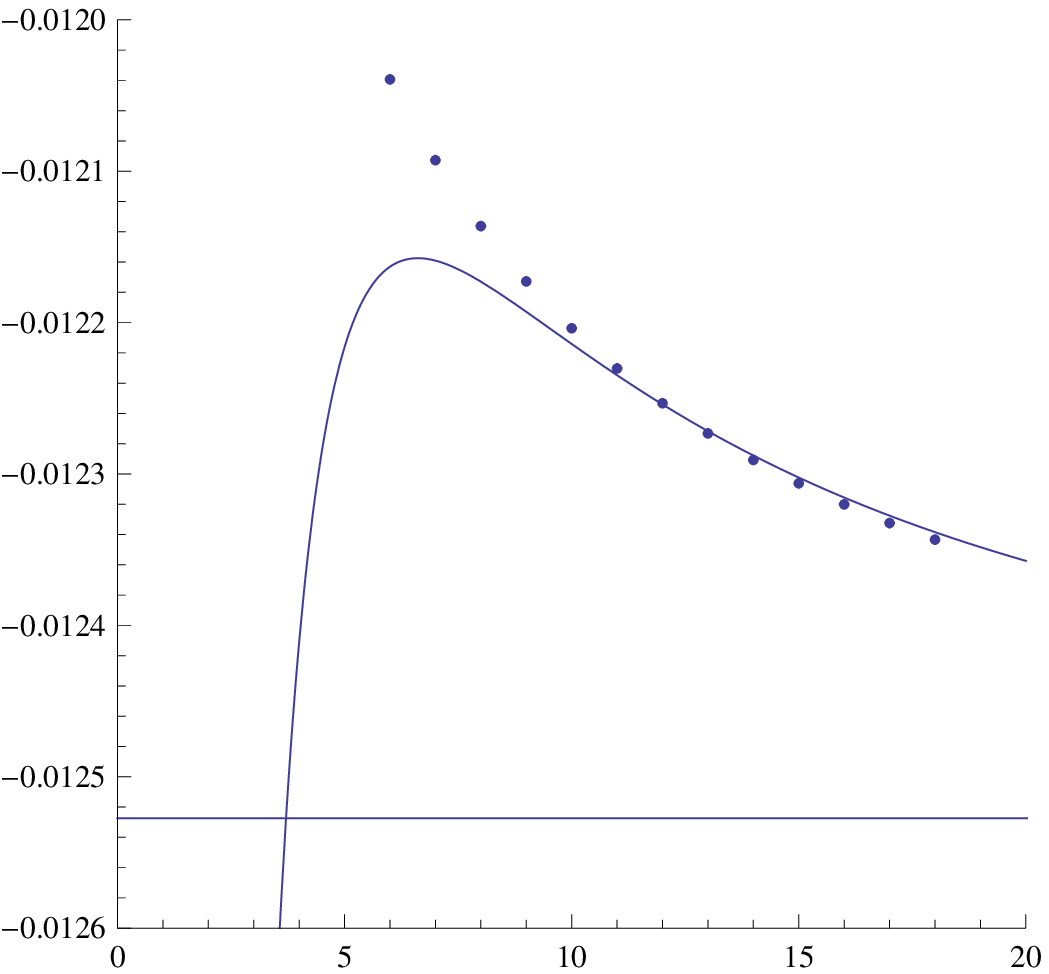}
\end{tabular}
\end{center}
\vspace{-5mm}
\caption{The coefficients of $\lambda^2$ in the first and second gap
 for the Yang-Lee model. The dots are from TCSA data, the solid curve
 the prediction using the two leading terms in $1/n$. The straight
 line is the exact value. } 
\label{fig:YLcoeffs}
\end{figure}

The Lee-Yang model two conformal boundary conditions which are
connected by an integrable boundary flow which has been studied in
great detail \cite{Dorey:1997yg}. 
The model is nonunitary and the boundary field generating the
boundary flow has conformal weight $-1/5$, which means that the
perturbation is UV finite but IR divergent.
Since $h=-1/5$, $y=1-h=6/5$ and consequently the leading corrections
from truncation errors, which are power series in $n^{-y}$, decay very
rapidly and the TCSA quickly becomes very accurate.
The truncation does still have an effect, of course, and the accuracy of
the numerical method allows us to check the predictions for the
coefficients of $\lambda^2$ in the first and second 
energy gaps. The results are shown in 
\ref{fig:YLcoeffs}, where we show the coefficients extracted from the
TCSA method and also the prediction based on the two leading terms in
$n^{-y}$. We see very good agreement so that the leading terms do
indeed give the correct behaviour. We have also checked this in the
tri-critical Ising model where we also get good agreement, but the
corrections from the terms in $n^{-3y}$ are larger.

We have used the following expression for the second energy level in
the Lee-Yang model, which is valid for any system where the
representation $\psi$ of conformal weight $H$ is degenerate at level 2
and has only a single 
state at that level which we have chosen to be $L_{-2}\vec\psi$:
\bea
  E_2
&=&
(H+2) + \tl C'\left(1 + \frac{4h(h{-}1)}{\Nc}\right)
\nonumber
\\
&+&\!\!\!\! (\tl C')^2\left[
  \frac{2h^2}{\Nc} {+} \frac{h^2(2h{-}1)^2}{2H\Nc} {-} 
\int_0^1 \frac{\rd z}{z^y}
\left( F^2 
  {-} \frac{4h^2}{\Nc z^{3-y}}
  {-} \frac{h^2(2h-1)^2}{(2H\Nc)z^{2-y}}
  {-} \frac{(4h(h{-}1){+}\Nc)^2}{(\Nc)^2z^{1-y}}
 \right)
\right]
\;,\;\;
\nonumber\\
 F^2
&=&
\frac{1}{\Nc}
\left(
   \left[ z(z^2-1)\frac{\rd}{\rd z} + (3z^2+1)h-2\right]
   \left[ (z-1/z)\frac{\rd}{\rd z} + h(3+1/z^2)\right]  
   + \Nc
\right) F^0
\;,\;\;
\nonumber\\
F^0 &=& 
\raisebox{-2.5mm}{\mbox{
$\cbb{2000}{\psi}{\phi}{\psi}{\phi}{\psi}{1}{z}
\;\;
$}}
\;,\;\;
\Nc = \frac{c}{2} + 4 H
\;,\;\;\;\;
\eea
where for the lee-Yang model $h=H=-1/5$ and
\be
F^0 = 
z^{1/5}(1-z)^{2/5} {}_2F_1(\tfrac 25,\tfrac 35;\tfrac 45;z)
\;,\;\;
C' = \left( \frac{\Gamma(1/5)\Gamma(2/5) }{\Gamma(-1/5)\Gamma(4/5) } 
     \right)^{1/2}
\;.
\ee

\newpage
\sect{Fixed points and the coupling constant renormalisation.} 
\label{eq:irfp}

\begin{figure}[thb]
\subfigure[The full TCSA spectrum showing the variation of the fixed point with
energy level; also shown is the best fit to these positions in the
scaling form
$\lambda_{\mathrm{crit}} = A (n-E)^y$, in the case $n=14$.
]{\scalebox{0.7}{\includegraphics{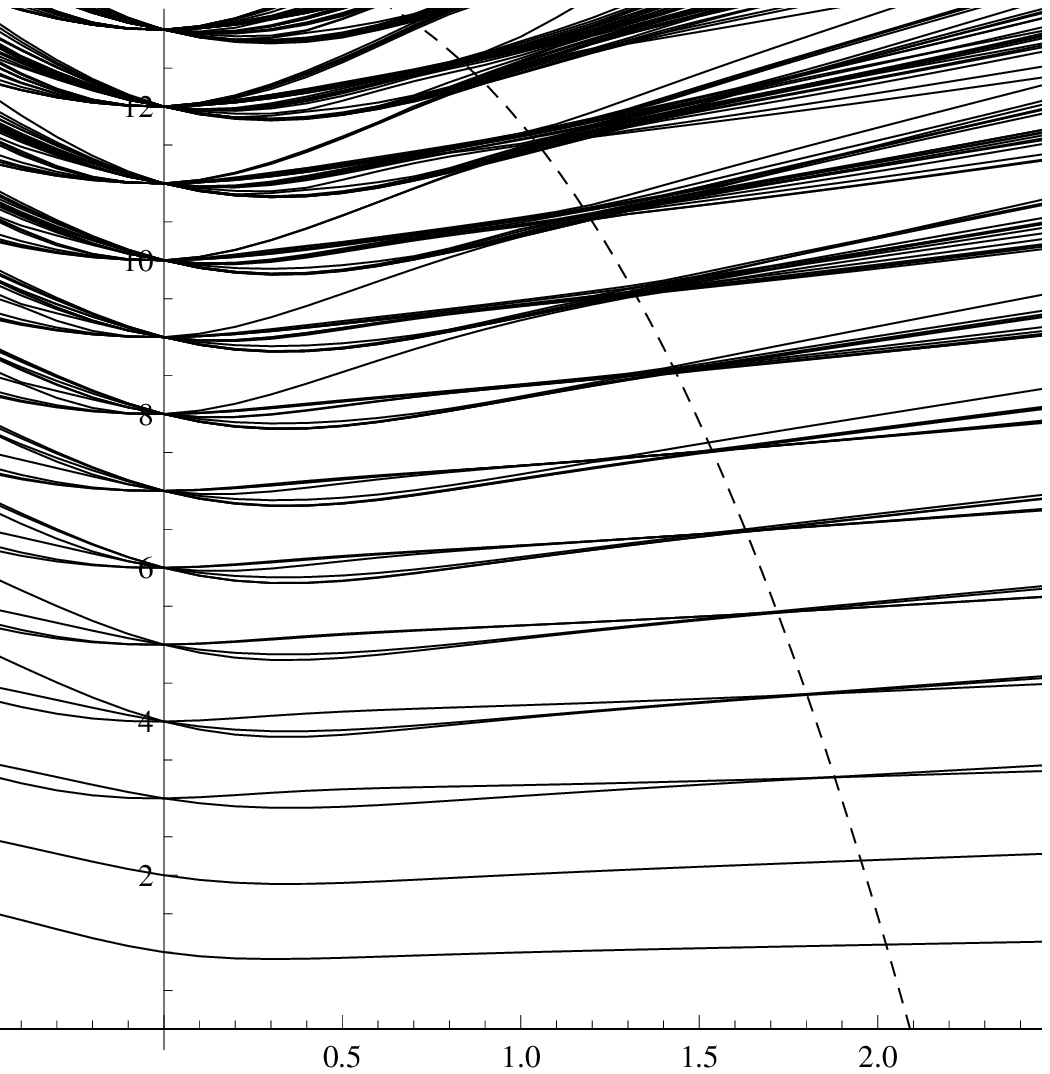}}%
\label{fig:ndepIR}}
\hfill
\subfigure[A simultaneous plot of the TCSA spectrum of the basic flow
  $(12)\to(21)$ on one edge and the fixed boundary condition
$(11)$ (red), $(21)$ (blue) and $(31)$ green on the other edge for
  truncation level 14.
]{\scalebox{0.7}{\includegraphics{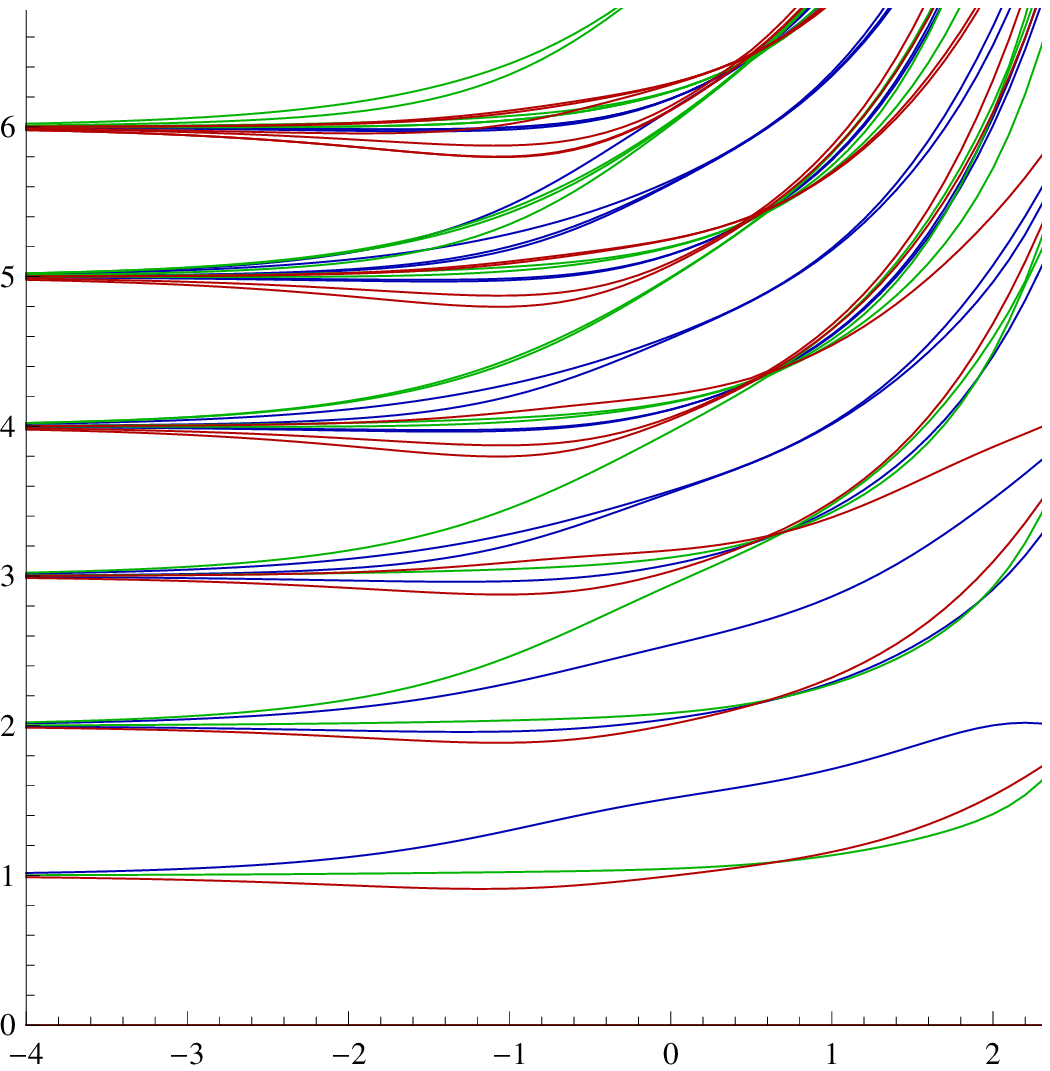}}
\label{fig:same}}
\caption{Energy-dependence and boundary-condition independence of the
  position of the fixed point in the flow $(12)\to(21)$.
}
\end{figure}

One of the principal effects of the coupling constant
renormalisation in the TCSA approach is the possibility to move the IR
fixed point of the finite-size scaling flow from infinite coupling
constant to finite coupling constant. 
From equations (\ref{eq:gndef}) and (\ref{eq:gn1loop}), we see that
for $C$ positive, the IR fixed point at 
$\lambda_\infty=\infty$ is brought to the finite value 
\be
\lambda_n^* = \frac{y\Gamma(h)}{C}(\pi n)^y
\;,
\label{eq:ndepIR}
\ee
However, just as the
coupling to the identity operator depends on the level of the state,
so the coupling constant renormalisation depends on the level of the
state, so that the one-loop effective coupling constant renormalisation of a
state at level $E$ is not given by (\ref{eq:gnform}), but rather by
\be
 \lambda_n^*(E)
= A(n) (n-E)^y
\;,
\label{eq:gnE}
\ee
where the one-loop prediction for $A(n)$ is $A(n) = y\Gamma(h)\pi^y/C$.
The exact position of the fixed point is not determined very
accurately by the 1--loop calculation, 
(indeed we shall see in section \ref{tcsa-compare} that the position of the
fixed point in the tri-critical Ising model is approximately
$\lambda^*_n = B \sqrt{n}$) 
but we might
conjecture is that the dependence on energy level still has this
form. In figure \ref{fig:ndepIR} we show the full TCSA spectrum for the
tri-critical Ising strip with boundary conditions $(12)$ and $(11)$. 
The level
crossings seen for positive $\lambda$ occur at the IR fixed point, and
their position depends on the excitation level. 
We have also shown a fit to these positions of the form
$\lambda_{\mathrm{FP}}(n,E) = A(n) (n-E)^y$.
This at least appears to capture the qualitative behaviour quite well.

A second prediction is that the position of this fixed point is
independent of the unperturbed boundary condition. In figure \ref{fig:same} 
we show the TCSA spectra of the three strips with one boundary being the
perturbation $(12)\to(21)$ and the other being one of the three fixed
boundary conditions $(r1)$. As can be seen, the three spectra agree
very well indeed on both the location of the fixed points, indicated
by the multiple line degeneracies, on the energy-dependence of these
fixed points and on the rescaling of the Hamiltonian, confirming the
prediction of our analysis that these effects are independent of the
unperturbed boundary condition (to leading order).

\sect{Flows generated by irrelevant fields}
\label{sec:irrel}

When we consider a flow with a UV and an IR fixed point, we can
attempt to describe it in terms of perturbations of either fixed
point. A perturbation of an IR fixed point will be by an irrelevant
field which is a non-renormalisable perturbation. This means that the
perturbation theory will have divergences that require regulation and
that a possibly infinite number of counterterms will be required to
allow one to remove the regulator. For this reason, they have only
infrequently been studied in the literature \cite{Zamo9,FQR1,Toth1}.
In TCSA, however, there is no need to introduce counterterms, one can
simply compute the spectrum of the truncated Hamiltonian. 

We have
investigated this in the case of the tri-critical Ising model.
There are three IR fixed points in the basic sequence \eref{eq:seq1}.
In the case of the $(11)$ and $(31)$ boundary conditions, the fields
on the boundary comprise the vacuum representation of the Virasoro
algebra so it is expected that the flow will be generated by the field
$T(x)$, the boundary stress-energy tensor, as in 
\cite{Toth1}; for the $(21)$ boundary condition, the boundary fields
form the $(11)$ and $(31)$ representations and it is to be expected
that the flow is generated by the field $\phi_{(31)}$ of weight $3/2$,
mirroring the case in the bulk flows considered in \cite{FQR1}.
We shall present some results for this last case, and for the flow
$(21)\to(13)$, in particular.

\subsection{Comparison with exact results}

\begin{figure}[thb] 
\subfigure[A plot of the effective coupling
$\log(\lambda^{(13)}_\infty)$ against 
$\log(\lambda^{(21)}_n)$. 
]{\scalebox{0.7}{\includegraphics{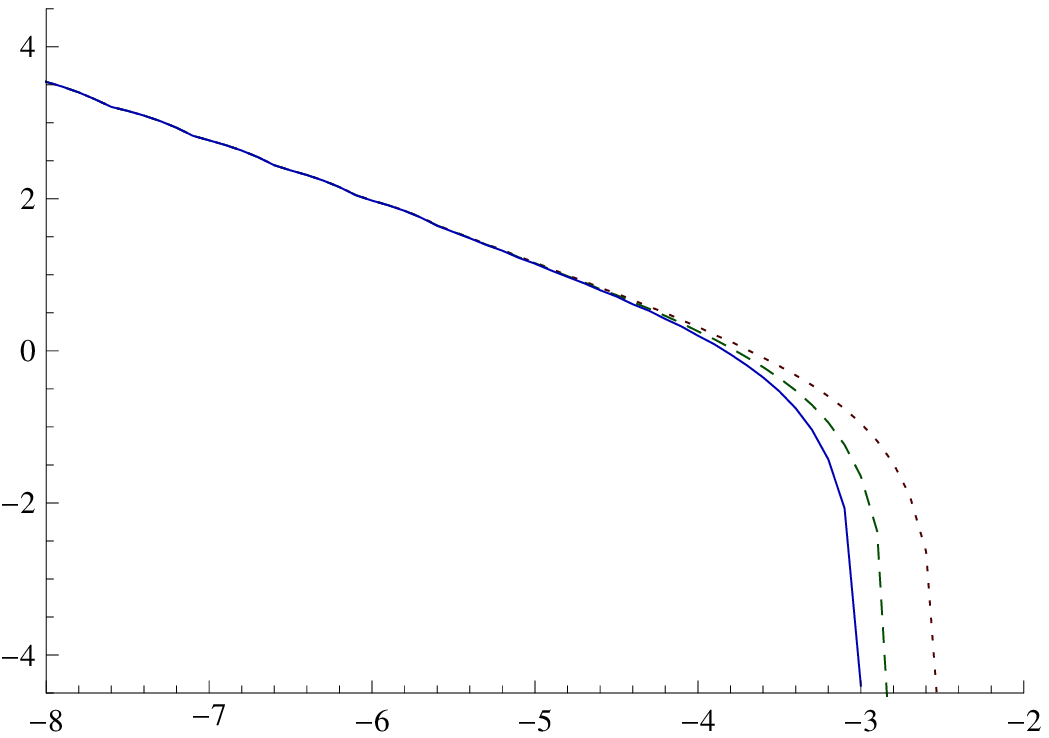}}%
\label{fig:31cc}}
\hfill
\subfigure[A plot of the scaling form 
$\lambda^{(13)}_\infty (\lambda^{(21)}_n)^{4/5}$ against
$\log(\lambda^{(21)}_n n^{1/2})$ together with the
limiting value of $-2.8$.
]{\scalebox{0.7}{\includegraphics{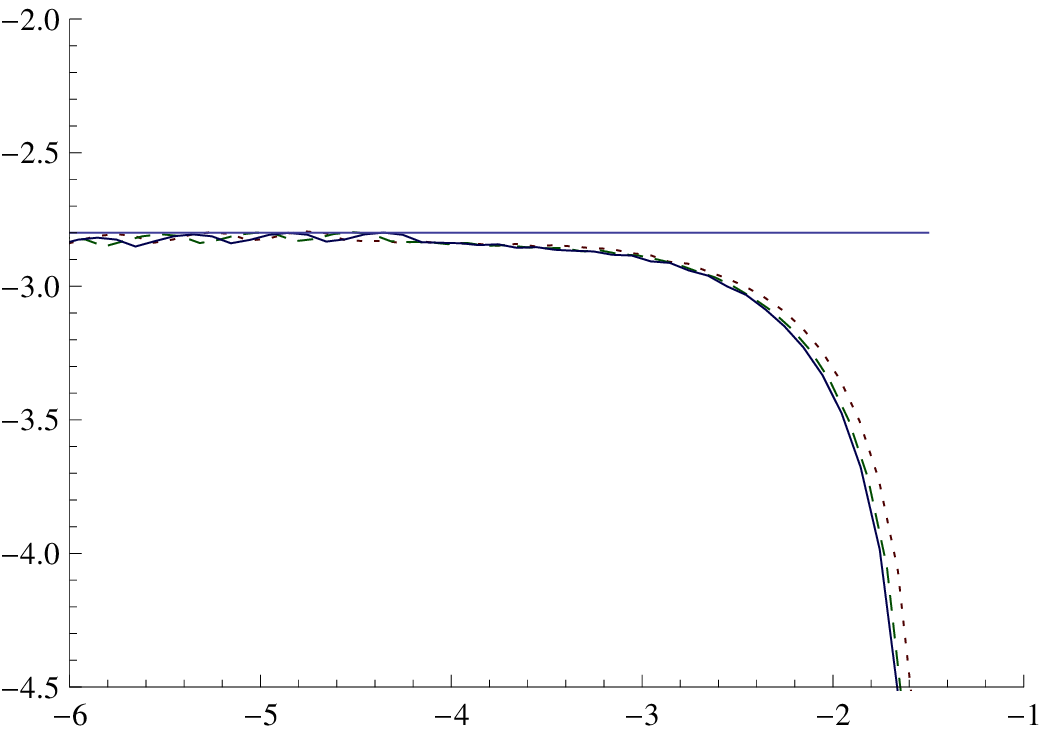}}
\label{fig:31ccg}}
\caption{Truncation level dependence and scaling functions for the
irrelevant coupling $\lambda^{(21)}$ at truncation levels 8 (red,
dotted), 13 (green, dashed) and 18 (blue, solid).
}
\label{fig:31ccs}
\end{figure}

Firstly, we find that the TCSA spectrum is in very good agreement with
the ``exact'' TBA spectrum;
we show the results for the spectrum in figure \ref{fig:overlap}.
Secondly, 
we can compare the TCSA couplings $\lambda^{(21)}_n$ at truncation
level $n$ with the effective coupling
$\lambda^{(13)}_\infty$ for the description of the system as a
perturbation of the $(13)$ by the field $\phi_{(13)}$. On dimensional
grounds, we expect a relation of the form
\be
  \lambda^{(13)}_\infty
= ( \lambda^{(21)}_n )^{-4/5} g( \lambda^{(21)}_n n^{1/2} )
\;,
\label{eq:31g}
\ee
and this indeed what we find. In figure \ref{fig:31cc} we plot the
effective couplings against the TCSA coupling for various truncation
levels, and in figure \ref{fig:31ccg} we plot the scaling function
$ \lambda^{(13)}_\infty (\lambda^{(21)}_n)^{4/5}$.

The scaling form \eref{eq:31g} does, however, have the major
consequence that the TCSA coupling constant does not converge to a
fixed value as the truncation level is increased. For an irrelevant
perturbation, $h>1$ and so $y<0$. If the function $g(x)$ in
\eref{eq:gnform} is approximately equal to one for a range $|x|<A$,
then the TCSA scaling function $g_n(\lambda)$ is approximately equal
to one for $|\lambda| < A n^{-y}$. For $y$ positive, this range grows
with increasing $n$; for an irrelevant perturbation it shrinks.
In this case, we see that
\be
\lambda^{(13)}_\infty (\lambda^{(21)}_n)^{4/5} \simeq 0.06
\;\;\;\; \hbox{for} \;\;\;\;
\lambda^{(21)}_n  < \frac{e^{-3}}{\sqrt n}
\;.
\label{eq:l1321exact}
\ee

\subsection{Comparison with TCSA parameters}
\label{tcsa-compare}

\begin{figure}[thb] 
\subfigure[%
A plot of 
$\lambda^{(13)}_n $ against
$\lambda^{(21)}_n $
]{\scalebox{0.7}{\includegraphics{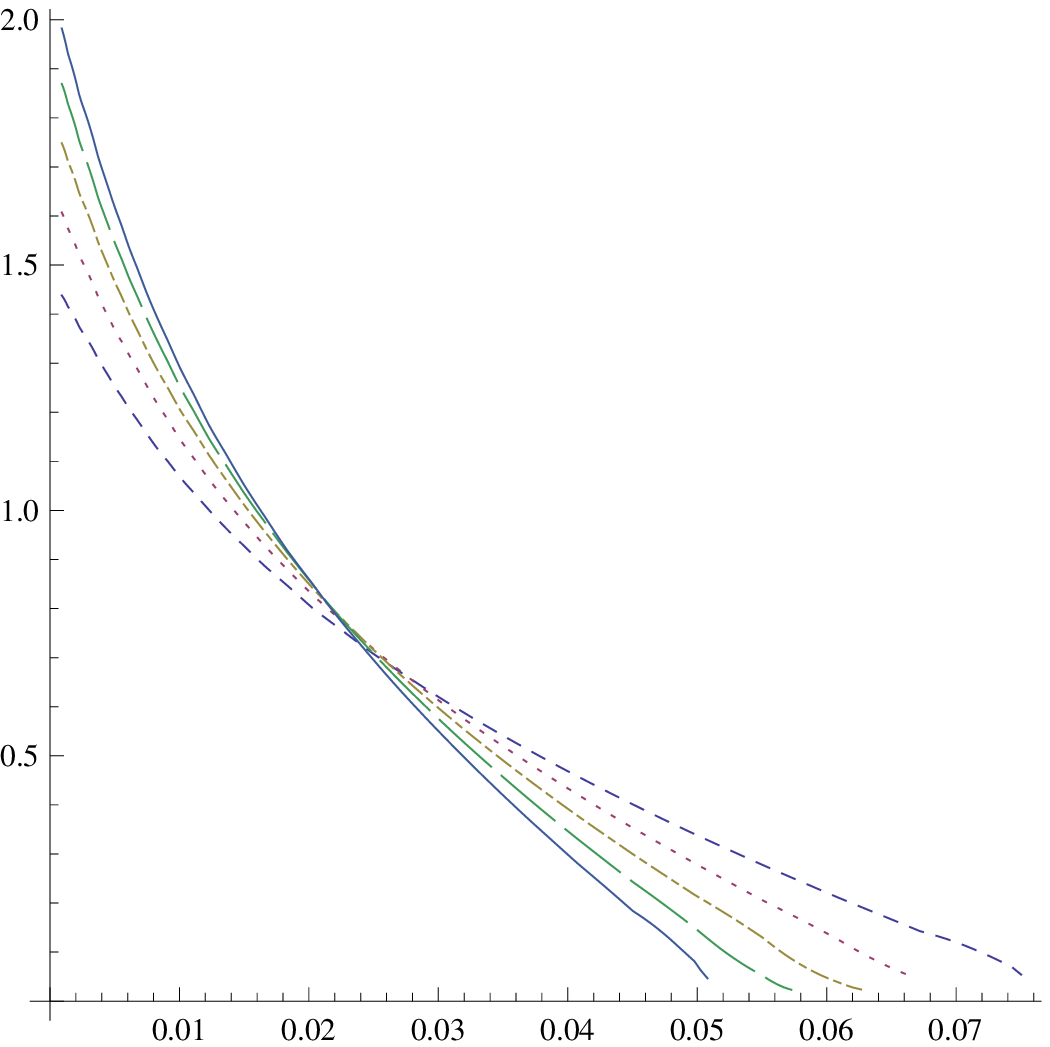}}
\label{fig:llplot1}}
\hfill
\subfigure[%
A plot of 
$\lambda^{(13)}_n / \sqrt n$ against
$\lambda^{(21)}_n \sqrt n)$
]{\scalebox{0.7}{\includegraphics{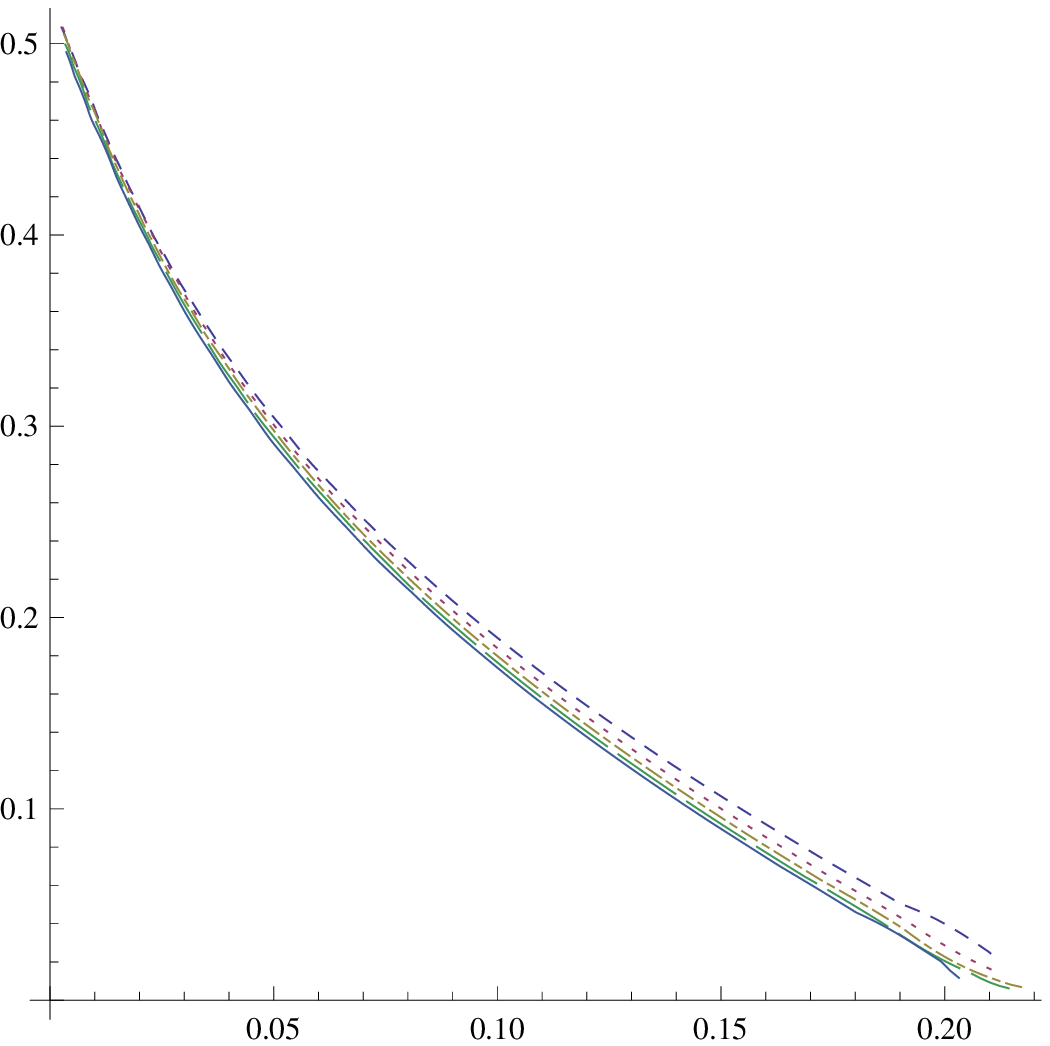}}
\label{fig:llplot2}}
\caption{The relation between $\lambda^{(13)}_n$ and
  $\lambda^{(21)}_n$ for truncation levels 8 (blue, short dashes), 10
  (dotted), 12 (dot-dashed), 14 (long dashes) and 22 (solid).
}
\end{figure}

While \eref{eq:31g} and \eref{eq:l1321exact} relate the TCSA parameter
$\lambda^{(21)}_n$ to the ``exact'' parameter $\lambda^{(13)}_\infty$,
it is also interesting to compare the two TCSA parameters. 
In this case we find the problem that the ranges in which the TCSA
parameters are good approximations to the ``exact'' parameters do not
overlap. 
The IR parameter is good for $\lambda^{(21)} < 0.05\, n^{-1/2}$, which
equates to $\lambda^{(13)} > .7 n^{2/5}$, 
while we can see from figure \ref{fig:gn} that 
the UV parameter is good for $\lambda^{(13)} < .5 n^{2/5}$ (where we
have used the $Z_2$ symmetry to relate $\lambda^{(13)}$ to
$-\lambda^{(12)}$.) 
As a consequence, there are no values of the TCSA coupling constants
for which \eref{eq:l1321exact} holds.
Instead we find the interesting plot \ref{fig:llplot1}; as $n$ increases,
the relation between the two TCSA couplings changes. For either of the
couplings small, there is an approximate linear relationship. We can
elucidate this by noting that the TCSA coupling relation itself takes
a scaling form as we show in figure \ref{fig:llplot2}, where we see
that, to a good approximation, 
\be
  \lambda^{(13)}_n \sqrt{n} \simeq F( \lambda^{(21)}_n / \sqrt{n} )
\;,
\label{eq:fx1}
\ee
where
\be
  F(x) \simeq 0.5 - 6. x
\; \hbox{for $x$ small,}
\qquad
  F(x) \simeq 0.3 - 1.3 x
\; \hbox{for $F(x)$ small}
\;.
\label{eq:fx}
\ee
The linear relation is to be expected, as the perturbing fields at one
end (UV or IR) will flow into the perturbing field at the other (IR or
UV, respectively) and so the coupling at one end will end up being
linearly related to the coupling at the other. Quite why this relation
takes the form \eref{eq:fx1} is unclear at the moment.

\sect{TCSA flows beyond the fixed point}
\label{sec:reverse}

\begin{figure}
\begin{center}
\scalebox{0.7}{
\includegraphics{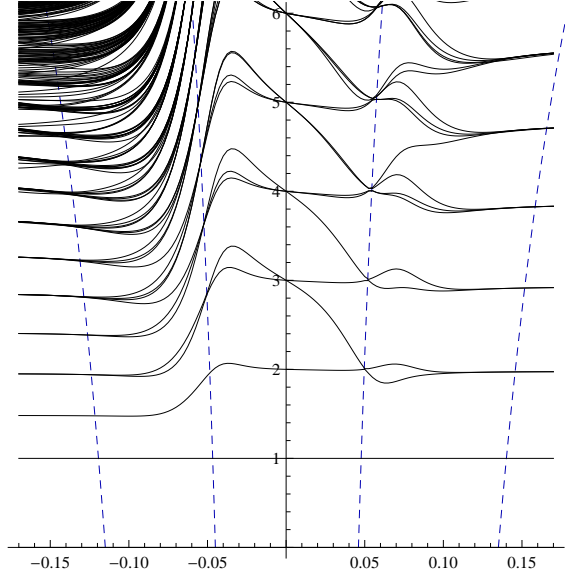}
}
\end{center}
\label{fig:21irplot}
\caption{The scaled gaps for the perturbation of the $(21)$ boundary
condition on a strip with boundary conditions $(21;11)$ by
$\lambda^{(21)}\phi_{(31)}$, plotted against $\lambda^{(21)}$ at
truncation level 14. The
four dashed lines are the approximate positions of the fixed points
(in order) $(11)$, $(12)$, $(13)$ and $(31)$.}
\end{figure}

As has been remarked on before \cite{FGPTW,Toth1}, the TCSA spectra have the
remarkable property that they extend beyond the IR fixed points and
can encompass several different perturbative flows. 
These appear to follow the sequence of flows found first by Lesage et
al in \cite{LSS1} and which also appear in Fredenhagen et al
\cite{Dorey1} and Dorey et al \cite{Dorey2} and
which in the simplest form applicable to 
the tri-critical Ising model is
\be
(12) \dashrightarrow
(11) \leftarrow
(12) \rightarrow
(21) \leftarrow
(13) \rightarrow
(31) \dashleftarrow
(13)
\label{eq:seq1}
\ee
where the dashed arrows reflect what is seen in TCSA and are a natural
extension of the sequence in the papers cited.
The same sequence is seen no matter which boundary condition we take
as the starting point from which we perturb. This applies also to the
case of irrelevant, non-renormalisable perturbations starting from the
`fixed' $(r1)$ boundary conditions. This enables us to put coordinates
on the sequence, in the sense that every pair of flows is covered as
the standard perturbation of a boundary condition, and we can relate
the coupling constants on successive overlapping pairs:
\be
\begin{array}{c@{}ccc@{}ccc@{}ccc@{}ccc@{}ccc@{}ccc@{}cccccc}
&&&
& \multicolumn{6}{c}{(12)+\mu^{(12)}\phi_{13}}
& \multicolumn{6}{c}{(13)+\mu^{(13)}\phi_{13}}
&\\
&&&
&\multicolumn{6}{c}%
{\overbrace{\phantom{1)\;\leftarrow\;(12)\;\rightarrow\;(2}}}
&\multicolumn{6}{c}%
{\overbrace{\phantom{1)\;\leftarrow\;(12)\;\rightarrow\;(2}}}
\\[-4mm]
  (1&2)
& \dashrightarrow
& (1&1)
& \leftarrow
& (1&2)
& \rightarrow
& (2&1)
& \leftarrow
& (1&3)
& \rightarrow
& (3&1)
& \dashleftarrow
& (1&3)
\\[-4mm]
&\multicolumn{6}{c}%
{\underbrace{\phantom{1)\;\leftarrow\;(12)\;\rightarrow\;(2}}}
&\multicolumn{6}{c}%
{\underbrace{\phantom{1)\;\leftarrow\;(12)\;\rightarrow\;(2}}}
&\multicolumn{6}{c}%
{\underbrace{\phantom{1)\;\leftarrow\;(12)\;\rightarrow\;(2}}}
\\[3mm]
&\multicolumn{6}{c}%
{(11)+\mu^{(11)}T+..}
&\multicolumn{6}{c}%
{(21)+\mu^{(21)}\phi_{31}+..}
&\multicolumn{6}{c}%
{(31)+\mu^{(31)}T+..}
\end{array}
\label{eq:seq2}
\ee
where the ellipsis in the perturbations of the fixed boundary
conditions shows that further counterterms may be needed as these
perturbations are non-renormalisable.
As an example of the way the TCSA flows can encompass several fixed
points, in figure \ref{fig:21irplot} we show the scaled gaps for the
perturbation of the $(21)$ boundary
conditions by the irrelevant field $\phi_{(31)}$ of conformal
weight $3/2$. As can be seen, all five fixed points of \eref{eq:seq1}
appear.

\begin{figure}
\begin{center}
\scalebox{0.7}{
\includegraphics{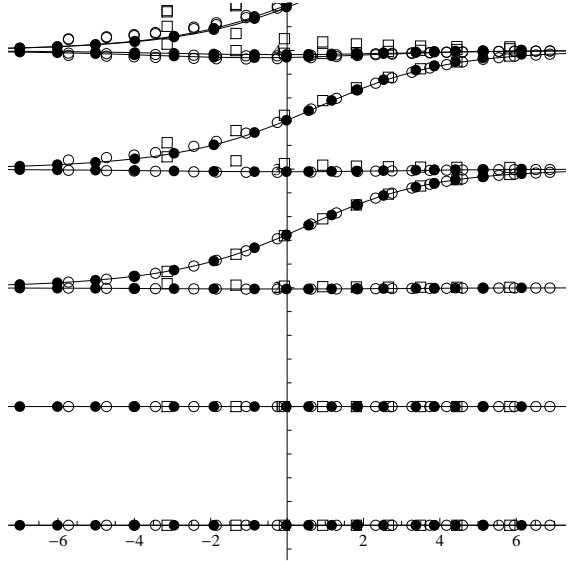}
}
\end{center}
\caption{The scaled gaps for the flow from the $(13)$ boundary to the
  $(21)$ boundary condition, as found from the relevant perturbation
  of the $(13)$ boundary condition ($\bullet$), from the irrelevant
  perturbation of the $(21)$ boundary condition  ($\circ$), and from
  the extension of the flow generated by the relevant perturbation of
  the $(12)$ boundary condition ($\square$) at truncation level
  22. They are plotted against $\xi$, the boundary parameter in Pearce
et al.}
\label{fig:overlap}
\end{figure}

To give support to this picture, we present the TCSA spectra for the
flow $(13)\to(21)$ obtained from the three separate routes: the 
relevant perturbation of the $(13)$ boundary condition, the irrelevant
perturbation of the $(21)$ boundary condition, and the continuation of
the flow $(12)\to(21)$ given by the relevant perturbation of the
$(12)$ boundary condition. These are all shown in figure
\ref{fig:overlap}. 

The TBA spectra are given as functions of the boundary reflection
parameter $\theta_B$ in \cite{NepomechieAhn}\footnote{The equations in
  \cite{LSS1,NepomechieAhn} for the $g$-function are known not to be
  correct, but they do describe the changes under a purely boundary
  perturbation. See \cite{Dorey2} for the correct equations when there
  is a simultaneous bulk perturbation.}%
or $\xi$ of Feverati et
al. \cite{FPR2} (note $\xi=\theta_B)$. Using the results in \cite{BLZ1,BLZ2}, 
we can find the exact expression for $\theta_B$ in terms of the
perturbative coupling constant $\lambda^{(13)}$ of the relevant
perturbation of the $(13)$ boundary condition.
We can also find the approximate numerical result for the
boundary parameter in terms of the TCSA coupling
$\lambda^{(21)}$, which is only valid for small coupling.
These are summarised below
\be
\renewcommand{\arraystretch}{1.6}
\begin{array}{c|ccl}
\hbox{Boundary}&&&\hbox{$\theta_B$}
\\ \hline
(13) &&&
\left(\tfrac 54 \log a -\log C_1\right) 
           + \tfrac 52 \log\lambda^{(13)}
\\
&&=& 1.96102... + \tfrac 52 \log\lambda^{(13)}
\\ \hline
(21) &&&
-5. - 2. \log\lambda^{(21)} + \ldots
\end{array}
\ee
The constants in the first relation are
\bea
    C_1 
=
    \tfrac 32 \pi \left[ \Gamma(\tfrac 15)\right]^{-5}
\;,\;\;\;\;
    a
=
    \frac{ \tfrac 35 \pi \sin(\tfrac 8 5 \pi)}
         {\Gamma( \tfrac 1 5 )^3 \Gamma(\tfrac 75 )
          \sin( \tfrac{12}{5}\pi)\sin(\tfrac{4}{5}\pi)}
\;.
\eea
Finally, we have also found an approximate numerical relation between
the couplings $\lambda^{(21)}$ and $\lambda^{(12)}$ for this region,
which is
\be
   \frac{\lambda^{(12)}_n}{\sqrt n}  - 8. \sqrt n \lambda^{(21)}_n   
\simeq  0.5
\;,
\label{eq:l2112}
\ee
which is in agreement with \eref{eq:fx} to the accuracy that is
obtainable for the flow from the $(12)$ boundary condition beyond the
$(21)$ fixed point.

\sect{Conclusions}
\label{sec:conc}

We have made further progress in understanding the errors in the
truncated conformal space approach to perturbed boundary conformal
field theory on a strip.
The principal quantity calculated with TCSA is the spectrum as a
function of the coupling constant.
The errors are of two principal sorts - a renormalisation of the
coupling constant and a change in the energy scale. We have found
perturbative expressions for these from physical arguments and showed
they are correct using an analysis of the perturbative spectrum.

One important aspect of these predictions is that they are
independent of the second boundary condition on the strip and we have
verified this by considering different choices for the unperturbed
boundary condition and finding the same results for each choice.

Furthermore we have investigated the behaviour of the TCSA spectrum
for large coupling constant. We have reported before that this appears
to show a sequence of RG flows in the same pattern as found by Lesage
et al \cite{LSS1}, and we know show that this is quantitatively
correct as well, in the case of the tri-critical Ising model.

It would be good to find some way to describe the full RG flows such
as in figure \ref{fig:21irplot} in terms of some beta-function which
has the sequence of fixed points as its zeroes, but at the moment we
are unsure how to do that. In the lattice model, the perturbation
parameter is a boundary magnetic field and the sequence of fixed
points \eref{eq:seq1} (excluding the points joined by the dashed
lines) can be found simply by varying this field \cite{Giokas2}; of
course in the quantum model this is not simply the case - the natural
sequence of flows splits up into overlapping pairs of flows in both the
TBA and the TCSA descriptions. 
It may be helpful to use the coordinates on
the full moduli space given by the values of the $g$-function and the
excited $g$-function which corresponds to the overlap with the bulk
spin field. This is something we hope to return to shortly,

Finally the ideas on the perturbative treatment of the corrections to
the TCSA presented here can be easily adapted to the case of
bulk flows which we plan to address in \cite{GW2}.

\sect{Acknowledgements}

I am very grateful to G.~Feverati, F.~Ravanini and P.A.~Pearce for
providing the numerical data for the TBA flows from their papers
\cite{FPR1,FPR2}, without which this paper would not have been possible.

I would like to thank P.~Giokas, G.~Tak\'acs and G.Zs.~T\'oth for helpful
discussions, B.~Doyon and G.~Tak\'acs for comments on the
manuscript and STFC grant 
ST/G000395/1 for support. 
All numerical calculations were performed using Mathematica \cite{mathematica}.

\label{sec:acc}

\newcommand\arxiv[2]      {\href{http://arXiv.org/abs/#1}{#2}}
\newcommand\doi[2]        {\href{http://dx.doi.org/#1}{#2}}
\newcommand\httpurl[2]    {\href{http://#1}{#2}}

\end{document}